\documentclass[fleqn,usenatbib]{mnras}

\usepackage{newtxtext,newtxmath}

\usepackage[T1]{fontenc}
\usepackage{ae,aecompl}

\usepackage{graphicx}	
\usepackage{amsmath}	

\usepackage{amssymb}	
\usepackage{longtable}
\usepackage{xcolor}
\usepackage{multirow}
\usepackage{pdflscape}
\usepackage{afterpage}

\title[H\,I Deficiencies and Asymmetries in HIPASS Galaxies]{H\,I Deficiencies and Asymmetries in HIPASS Galaxies.}

\author[T.N.~Reynolds, et al.]{T.N.~Reynolds$^{1,2}$\thanks{tristan.reynolds@uwa.edu.au}, 
T.~Westmeier$^{1,2}$,
L.~Staveley-Smith$^{1,2}$
\\
$^1$International Centre for Radio Astronomy Research (ICRAR), The University of Western Australia, 35 Stirling Hwy, Crawley, WA, 6009, Australia\\
$^2$ARC Centre of Excellence for All Sky Astrophysics in 3 Dimensions (ASTRO 3D)
}

\date{Accepted 2020 October 06. Received 2020 September 29; in original form 2020 September 03}

\pubyear{2020}

\begin{document}
\label{firstpage}
\pagerange{\pageref{firstpage}--\pageref{lastpage}}
\maketitle

\begin{abstract}
We present an analysis of the sky distribution of neutral hydrogen (H\,\textsc{i}) deficiency and spectral asymmetry for galaxies detected by the H\,\textsc{i} Parkes All-Sky Survey (HIPASS) as a function of projected environment density. Previous studies of galaxy H\,\textsc{i} deficiency using HIPASS were sensitive to galaxies that are extremely H\,\textsc{i} rich or poor. We use an updated binning statistic for measuring the global sky distribution of H\,\textsc{i} deficiency that is sensitive to the average deficiencies. Our analysis confirms the result from previous studies that galaxies residing in denser environments, such as Virgo, are on average more H\,\textsc{i} deficient than galaxies at lower densities. However, many other individual groups and clusters are not found to be on average significantly H\,\textsc{i} poor, in contradiction to previous work. In terms of H\,\textsc{i} spectral asymmetries, we do not recover any significant trend of increasing asymmetry with environment density as found for other galaxy samples. We also investigate the correlation between H\,\textsc{i} asymmetry and deficiency, but find no variation in the mean asymmetry of galaxies that are H\,\textsc{i} rich, normal or poor. This indicates that there is either no dependence of asymmetry on H\,\textsc{i} deficiency, or a galaxy's H\,\textsc{i} deficiency only has a small influence on the measured H\,\textsc{i} asymmetry that we are unable to observe using only integrated spectra.
\end{abstract}




\section{INTRODUCTION}
\label{sec:intro}

Galaxies are observed with morphologies spanning from active star forming, late-type systems through to passive, early-types. Across morphologies, asymmetries are observed in the stellar and gaseous (atomic hydrogen, H\,\textsc{i}) disks, and galaxies are found with gas contents ranging from gas-rich to gas-poor. A late-type, star forming galaxy in isolation and without any external influences is most likely to appear symmetrical as the morphology is primarily driven by the gravitational potential well arising from the baryonic (stellar, gas) and non-baryonic (dark) matter. 

Most galaxies do not live in isolation, but are found to reside in groups and clusters \citep[e.g.][]{Tully1987,Gourgoulhon1992} and large fractions of galaxies are observed with measurable asymmetries in their integrated H\,\textsc{i} spectra \citep[e.g.][]{Richter1994,Haynes1998,Matthews1998}. The environment is known to be important in shaping the observed galaxy properties \citep[e.g.\ the morphology-density relation,][]{Dressler1980}. One such property is a galaxy's H\,\textsc{i} deficiency, which is defined as the difference between the observed H\,\textsc{i} mass and the expected H\,\textsc{i} mass for an isolated galaxy of similar size and morphology \citep[e.g.][]{Haynes1984,Cortese2011}. H\,\textsc{i} deficient galaxies are more frequently observed in denser environments, particularly in galaxy clusters (e.g.\ the Virgo cluster) and towards the centre of galaxy groups, rather than in isolated, field galaxies and galaxies in low density groups \citep[e.g.][]{Chamaraux1980,Solanes2001,Verdes-Montenegro2001,Boselli2009,Chung2009,Kilborn2009,Hess2013}.

There are many external (e.g.\ tidal and ram pressure stripping) and internal (e.g.\ star formation and supernova feedback) mechanisms proposed for causing the diversity of galaxy morphologies and giving rise to observed asymmetries, with the occurrence of each mechanism depending on the environment in which the galaxy resides (e.g.\ isolated in the field, intermediate densities in groups, or high densities in clusters). External mechanisms fall under two broad categories: mechanisms acting between a galaxy and the environment and mechanisms acting between multiple galaxies. Mechanisms involving the environment include ram pressure stripping \citep{Gunn1972} as a galaxy passes through the dense intergalactic medium (IGM) and gas accreting asymmetrically onto a galaxy \citep{Bournaud2005b}. Interactions between galaxies with low relative velocities \citep[tidal stripping,][]{Moore1999,Koribalski2009,English2010} and high relative velocities \citep[`harassment',][]{Moore1996,Moore1998} and galaxy mergers \citep{Zaritsky1997}, on the other hand, involve two or more galaxies. H\,\textsc{i} deficient galaxies in groups and clusters are also found to be undergoing ram pressure stripping or tidal interactions \citep[e.g.][]{BravoAlfaro2000,Denes2014,Denes2016,Yoon2017}, which indicates that these external mechanisms are likely to be the cause of the observed H\,\textsc{i} deficiencies.

The influence of the environment can be very effectively probed with H\,\textsc{i}, which will be affected by external mechanisms more readily than the stellar disk \citep[e.g.][]{Giovanelli1985,Solanes2001,Rasmussen2006,Rasmussen2012,Westmeier2011,Denes2014,Odekon2016}. H\,\textsc{i} is subject to pressure effects and is easily observed to larger radii than the stellar disk, which is where the gravitational potential is weaker and the gas will be less gravitationally bound. The most straightforward observable relevant for galaxy asymmetry is a galaxy's integrated 21\,cm spectrum, which can be acquired with either a single dish radio telescope or an interferometer and requires significantly less observing time compared with spatially resolved observations. For these reasons, many H\,\textsc{i} asymmetry studies have focused on asymmetries in integrated spectra. 

The most common parameter for quantifying the level of asymmetry is the flux ratio asymmetry, $A_{\mathrm{flux}}$, defined by taking the ratio of the integrated flux above and below the systemic velocity of a galaxy. Early studies found $\gtrsim50\%$ of galaxies to be asymmetric with flux ratios $>1.05$ \citep[][]{Richter1994,Haynes1998,Matthews1998}. The scatter in the flux ratio asymmetry of the AMIGA \citep[Analysis of the interstellar Medium in Isolated GAlaxies,][]{VerdesMontenegro2005} isolated galaxy sample follows a half Gaussian centred on 1, denoting a completely symmetric system, with a standard deviation of 0.13 \citep{Espada2011}. Hence, the fraction of systems classified as asymmetric in previous studies will be lower if the asymmetry cutoff adopted is based on the isolated galaxy sample of \cite{Espada2011}. More recent studies by \cite{Scott2018} and \cite{Bok2019} have considered galaxies with $A_{\mathrm{flux}}>1.26$ or 1.39 (i.e.\ $>2\sigma$ or $3\sigma$) as being significantly asymmetric and the asymmetry likely caused by external mechanisms. Using $>2\sigma$, this corresponds to $9\%$, $17\%$, $2\%$ and $18\%$ for the isolated galaxy samples (\citeauthor{Haynes1998}, \citeyear{Haynes1998}; \citeauthor{Matthews1998}, \citeyear{Matthews1998}; \citeauthor{Espada2011}, \citeyear{Espada2011}; \citeauthor{Bok2019}, \citeyear{Bok2019}, respectively), $\sim16$--$26\%$ for the Virgo and Abell\,1367 clusters \citep{Scott2018} and $27\%$ for galaxy pairs \citep{Bok2019}. 

Studies of H\,\textsc{i} asymmetries are not limited to integrated spectra, but have also quantified asymmetries in H\,\textsc{i} morphologies and kinematics \citep[e.g.][]{Angiras2006,Angiras2007,vanEymeren2011a,vanEymeren2011b,Reynolds2020}. However, measuring morphological and kinematic asymmetries requires spatially resolved, interferometric observations, which require significant amounts of observing time. There are currently no all-sky H\,\textsc{i} surveys of spatially resolved galaxies similar to the H\,\textsc{i} Parkes All-Sky Survey \citep[HIPASS,][]{Barnes2001}, which is limited to integrated spectra.

Quantifying the environmental dependence of any galaxy property is complicated by how the environment is defined \citep[e.g.\ cross matching using a group catalogue or optical survey, optical survey magnitude limits, nearest neighbour vs fixed aperture, number of nearest neighbours, volume vs projected density, etc.,][]{Muldrew2012, Jones2016}. Some of the effects of these variables can be mitigated with a sufficiently large sky coverage and complementary optical sky coverage. \cite{Denes2014} used HIPASS and HyperLEDA \citep{Paturel2003,Makarov2014} to map the sky distribution of H\,\textsc{i} deficiency compared with the large-scale galaxy structure finding H\,\textsc{i} deficient areas on the sky which are correlated with the locations of groups and clusters. We can apply the same analysis technique to the integrated H\,\textsc{i} flux asymmetry ratio of HIPASS galaxies to investigate the global distribution of H\,\textsc{i} asymmetries across the southern sky. 

It is not yet possible to map the sky distribution of H\,\textsc{i} asymmetries on the basis of spatially resolved observations. However, this will change in the next few years with the Widefield ASKAP L-band Legacy All-sky Blind Survey \citep[WALLABY,][]{Koribalski2020} which will be carried out on the Australian Square Kilometre Array Pathfinder \citep[ASKAP,][]{Johnston2007}. WALLABY will cover $\sim75\%$ of the sky up to $\delta<+30^{\circ}$ out to $z<0.26$ and will spatially resolve all HIPASS sources (WALLABY and HIPASS synthesised beams: $0.5\arcmin$ vs $15.5\arcmin$, respectively). WALLABY is predicted to detect $\sim500\,000$ galaxies of which several thousand will be spatially resolved while the rest will be limited to integrated spectra. With an expected mean redshift of $z\sim0.05$, WALLABY will be able to map spectral asymmetry sky distributions out to much higher redshifts than possible with HIPASS (mean redshift $z\sim0.01$).

In this work, we investigate the distribution of H\,\textsc{i} flux asymmetry ratio, $A_{\mathrm{flux}}$, in HIPASS galaxies across the sky and examine the correlation with large-scale structure. We also carry out a new analysis of the sky distribution of H\,\textsc{i} deficiency similar to \cite{Denes2014} where we measure the mean rather than the summed H\,\textsc{i} deficiency across the sky. Finally, we compare the global H\,\textsc{i} asymmetry and H\,\textsc{i} deficiency. We describe the data we use and our analysis in Sections~\ref{sec:data} and \ref{sec:analysis}, respectively. We discuss our results in Section~\ref{sec:discussion} and present our conclusions in Section~\ref{sec:conclusion}. Throughout, we use velocities in the optical convention (c$z$) and the Local Group reference frame, adopting a flat $\Lambda$CDM cosmology using $H_0=67.7$\,km\,s$^{-1}$\,Mpc$^{-1}$, concordant with \textit{Planck} \citep{Planck2016}.

\section{DATA}
\label{sec:data}

\subsection{HIPASS}
\label{s-sec:hipass}

The H\,\textsc{i} Parkes All Sky Survey \citep[HIPASS,][]{Barnes2001} is a blind H\,\textsc{i} survey covering the sky at $\delta<+25^{\circ}$ carried out on the Parkes 64-m telescope with a gridded beam size of $15.5\arcmin$ and a spectral resolution of 18\,km\,s$^{-1}$ (with a channel width of $\sim13.2$\,km\,s$^{-1}$). Two H\,\textsc{i} catalogues were produced from HIPASS covering the southern and northern portions of the sky \citep[HICAT and NHICAT,][respectively]{Meyer2004,Wong2006}. HICAT contains 4\,315 sources at $\delta<+2^{\circ}$ and $v>300$\,km\,s$^{-1}$ with optical identifications and properties catalogued in HOPCAT \citep{Doyle2005}. NHICAT covers the northern extension of HIPASS, $+2^{\circ}<\delta<+25^{\circ}$, and contains 1\,002 sources at $v>300$\,km\,s$^{-1}$ with optical identifications and properties catalogued in NOIRCAT \citep{Wong2009}. We summarise the basic properties of the two catalogues in Table~\ref{table:hipass_obs_params}. The lower velocity limit of $v>300$\,km\,s$^{-1}$ ensures there is no contamination by Galactic emission in the detected sources.

We extract all HIPASS spectra listed in HICAT and NHICAT following the method of \cite{Meyer2004} and \cite{Wong2006}, which we summarise here. The HICAT and NHICAT catalogues include the position of each galaxy, the number of pixels on a side of the box used to extract each source and whether sources are extended or point sources. For each source, we use the \textsc{miriad} task \textsc{mbspect} to extract a box (size listed in HICAT and NHICAT) centred on the pixel containing the HIPASS source position covering all channels with $v>300$\,km\,s$^{-1}$. The pixels in each channel of the extracted box are weighted by the gridded HIPASS beam (Gaussian with $\mathrm{FWHM}=15.5\arcmin$) for HIPASS classified `point' sources and are summed and normalised for `extended' sources. We note that the actual size of the HIPASS beam is dependent on the source peak flux density and size and varies by $\pm1'$ \citep{Barnes2001}. However, we are only interested in the relative flux in each half of the spectra for calculating the flux asymmetry (Section~\ref{s-sec:asyms}) in which case the scaling factor from weighting by the Gaussian beam cancels out. \cite{Meyer2004} and \cite{Wong2006} use the accurate beam sizes for their measured integrated H\,\textsc{i} fluxes in HICAT and NHICAT, respectively, which we use for calculating the H\,\textsc{i} deficiency. We remove any residual baseline offsets in the extracted spectra by fitting and subtracting a $2^{\mathrm{nd}}$-order polynomial to the spectra after masking out the velocity range containing the galaxy, which is tabulated in HICAT and NHICAT. We finally mask channels outside of the velocity range containing each galaxy from HICAT and NHICAT before measuring the asymmetry as described in Section~\ref{s-sec:asyms}.

\begin{table}
	\centering
    \caption{HIPASS source catalogues.}
	\label{table:hipass_obs_params}
	\begin{tabular}{lcr}
		\hline
		 & HICAT & NHICAT \\ \hline
		Region & South & North  \\
		Galaxies [$N$] & 4\,315 & 1\,002 \\
		Optical & HOPCAT & NOIRCAT \\
		RA & $0^{\circ}<\alpha<360^{\circ}$ & $0^{\circ}<\alpha<360^{\circ}$ \\
        Dec & $-90^{\circ}<\delta<+2^{\circ}$ & $+2^{\circ}<\delta<+25^{\circ}$ \\ \hline
	\end{tabular}
\end{table}

\subsection{HyperLEDA}
\label{s-sec:hyperleda}

Following \cite{Denes2014}, as a reference catalogue of the sky distribution of galaxies we use HyperLEDA\footnote{\url{http://leda.univ-lyon1.fr/}} \citep{Paturel2003,Makarov2014}, which provides mean homogenised galaxy parameters. We include all HyperLEDA galaxies down to the apparent magnitude completeness limit of $m_B=14$\,mag to which HyperLEDA is complete \citep{Giuricin2000}. Table~\ref{table:absolute_magnitudes} lists the absolute $B-$band magnitude limits at the velocity bounds for the four velocity ranges we use to bin galaxies: $V_1=300$--1\,000\,km\,s$^{-1}$, $V_2=1\,000$--2\,000\,km\,s$^{-1}$, $V_3=2\,000$--4\,000\,km\,s$^{-1}$ and $V_4=4\,000$--6\,000\,km\,s$^{-1}$. We note that the difference in the absolute magnitude limits covers 6 orders of magnitude between the lowest and highest velocities. This limits our analysis of the variation of asymmetries and deficiencies with environment density to within each velocity range and we do not draw conclusions about variation between velocity bins. 

\begin{table}
	\centering
    \caption{HyperLEDA absolute $B-$band magnitude limits at the lower and upper velocity bounds for the four velocity ranges: $V_1=300$--1\,000\,km\,s$^{-1}$, $V_2=1\,000$--2\,000\,km\,s$^{-1}$, $V_3=2\,000$--4\,000\,km\,s$^{-1}$ and $V_4=4\,000$--6\,000\,km\,s$^{-1}$.}
	\label{table:absolute_magnitudes}
	\begin{tabular}{lr}
		\hline
		Velocity & $M_B$ \\
		{[km\,s$^{-1}$]} & [mag] \\ \hline
		300 & $-14.23$ \\
		1\,000 & $-16.85$ \\
		2\,000 & $-18.35$ \\
		4\,000 & $-19.86$ \\
		6\,000 & $-20.74$ \\ \hline
	\end{tabular}
\end{table}

\section{ANALYSIS}
\label{sec:analysis}

\subsection{HI Deficiency}
\label{s-sec:hi_def}

We perform a similar analysis to that of \cite{Denes2014} for the spatial distribution of H\,\textsc{i} deficiency across the sky for comparison with the asymmetry sky distributions. For consistency, we use the same subsample of HIPASS sources defined by \cite{Denes2014}, which only includes sources with single optical counterparts and integrated fluxes $S_{\mathrm{int}}>5$\,Jy\,km\,s$^{-1}$ \citep[i.e.\ the HIPASS 95$\%$ reliability limit from][]{Zwaan2004}. We compute the expected and observed H\,\textsc{i} masses using equations~4 and 5 of \cite{Denes2014}, respectively. The expected H\,\textsc{i} masses are calculated using the scaling relations derived from the $B-$band for HICAT and NHICAT \citep[tables~2 and 3 in][respectively]{Denes2014} and using the $B-$band magnitudes from HOPCAT and NOIRCAT. The H\,\textsc{i} deficiency, $\mathrm{DEF}_{\mathrm{HI}}$ is calculated in the normal way,
\begin{equation}
	\displaystyle \mathrm{DEF}_{\mathrm{HI}}=\log_{10}(M_{\mathrm{exp}}/\mathrm{M}_{\odot})-\log_{10}(M_{\mathrm{obs}}/\mathrm{M}_{\odot}),
	\label{equ:hi_def}
\end{equation}
where $M_{\mathrm{exp}}$ and $M_{\mathrm{obs}}$ are the expected and observed H\,\textsc{i} masses, respectively. Galaxies are considered to be H\,\textsc{i} deficient if $\mathrm{DEF}_{\mathrm{HI}}>0.3$ and H\,\textsc{i} rich if $\mathrm{DEF}_{\mathrm{HI}}<-0.3$ (i.e.\ compared to an average spiral galaxy either double or half the quantity of H\,\textsc{i}, respectively). The global H\,\textsc{i} deficiency map is presented in Section~\ref{s-sec:global_def}.

\subsection{HI Asymmetry}
\label{s-sec:asyms}

\begin{figure}
	\centering
	\includegraphics[width=\columnwidth]{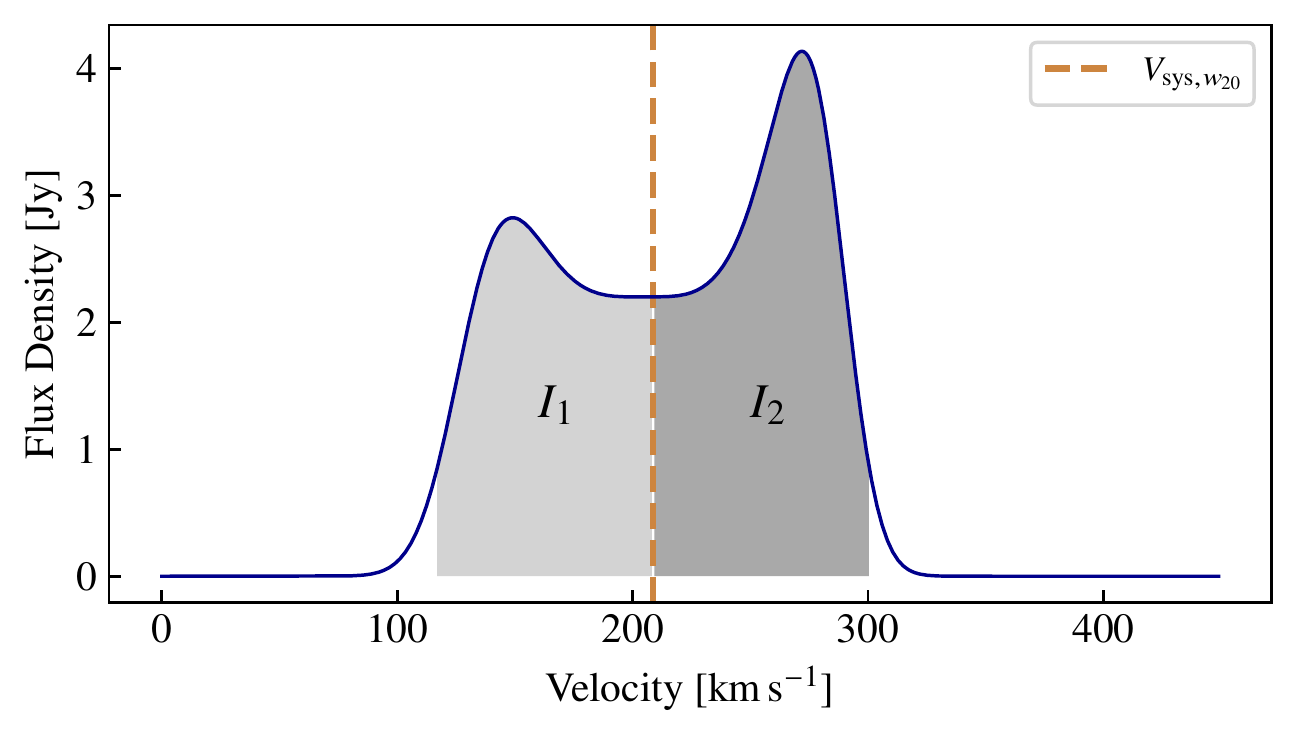}
		\caption{Example of an asymmetric H\,\textsc{i} spectrum indicating the systemic velocity defined as the midpoint of the spectrum at $20\%$ the spectrum's peak height (dashed orange line). Also indicated are the light and dark shaded regions integrated to calculate $A_{\mathrm{flux}}$ ($I_1$ and $I_2$, respectively). For this spectrum $A_{\mathrm{flux}}=1.09$.}
	\label{fig:example_spectrum}
\end{figure}

We quantify the spectral H\,\textsc{i} asymmetry with the flux ratio asymmetry, $A_{\mathrm{flux}}$, which is frequently used to quantify the asymmetry in an integrated H\,\textsc{i} spectrum \citep[e.g.][]{Richter1994,Haynes1998,Espada2011,Scott2018}. $A_{\mathrm{flux}}$ is calculated by taking the ratio of the integrated flux in the two halves of the spectrum split at the source's systemic velocity,
\begin{equation}
	\displaystyle A_{\mathrm{flux}}=\frac{I_1}{I_2}=\frac{\sum_{v_{\mathrm{low}}}^{V_{\mathrm{sys},w_{20}}} Idv}{\sum_{V_{\mathrm{sys},w_{20}}}^{v_{\mathrm{high}}} Idv},
	\label{equ:hipass_flux_asym}
\end{equation}
where $I_{1}$ and $I_{2}$ are the integrated fluxes in the lower and upper halves of the spectrum (shaded regions in Figure~\ref{fig:example_spectrum}). The systemic velocity, $V_{\mathrm{sys},w_{20}}$, is defined as the midpoint of the spectrum at 20$\%$ of the peak flux (the $w_{20}$ line width) and $v_{\mathrm{low}}=V_{\mathrm{sys},w_{20}}-w_{20}/2$ and $v_{\mathrm{high}}=V_{\mathrm{sys},w_{20}}+w_{20}/2$. We follow \cite{Reynolds2020} and interpolate at the upper and lower boundaries of the two regions $I_{1}$ and $I_{2}$. For channels bridging the boundaries of $I_{1}$ and $I_{2}$, we divide the flux in the bridging channel proportionally between $I_{1}$ or $I_{2}$ by the fractional channel width that lies within either half of the spectrum. The flux asymmetry ratio is defined to be $A_{\mathrm{flux}}\geq1$ (i.e.\ if $A_{\mathrm{flux}}=I_1/I_2<1$ then we take the inverse: $A_{\mathrm{flux}}=I_2/I_1$), where $A_{\mathrm{flux}}=1$ for a symmetric spectrum and $A_{\mathrm{flux}}$ increases with increasing asymmetries. The global flux asymmetry ratio map is presented in Section~\ref{s-sec:global_asym}. In Appendix~\ref{appendix:sample_spec} we show 20 example HIPASS spectra with $V_{\mathrm{sys},w_{20}}$, $I_1$ and $I_2$ indicated as in Figure~\ref{fig:example_spectrum} along with each spectrum's $A_{\mathrm{flux}}$ (Figure~\ref{fig:hipass_example_spectra}).

If a galaxy's $w_{20}$ line width is comparable to the channel width (e.g.\ the spectrum is resolved by $\lesssim10$ channels, $\sim7$ spectral resolution elements) then the measured $A_{\mathrm{flux}}$ can be blurred/diluted due to instrument resolution. This does not affect our sample as there are only 6 galaxies (0.5\%) with $\mathrm{SNR}>30$ that have $w_{20}<50$\,km\,s$^{-1}$ (86 galaxies, 7\%, with $w_{20}<100$\,km\,s$^{-1}$).

\begin{table}
	\centering
    \caption{Total HIPASS spectra above integrated signal to noise ratio (SNR) cuts and subsamples in each of the four velocity ranges used in sky plots: $V_1=300$--1\,000\,km\,s$^{-1}$, $V_2=1\,000$--2\,000\,km\,s$^{-1}$, $V_3=2\,000$--4\,000\,km\,s$^{-1}$ and $V_4=4\,000$--6\,000\,km\,s$^{-1}$.}
	\label{table:snr_cuts}
	\begin{tabular}{lccccr}
		\hline
		SNR & \multicolumn{5}{c}{Galaxies [$N$]} \\
		 & All & $V_1$ & $V_2$ & $V_3$ & $V_4$ \\ \hline
		10 & 4\,617 & 426 & 1\,205 & 1\,734 & 889 \\
		20 & 2\,137 & 294 &  730 &  791 & 250 \\
		30 & 1\,167 & 221 &  474 &  384 &  69 \\
		40 &  744 & 180 &  329 &  204 &  21 \\ \hline
	\end{tabular}
\end{table}

$A_{\mathrm{flux}}$ can be calculated for any integrated H\,\textsc{i} spectrum with a sufficiently high integrated signal to noise ratio (SNR) and spectral resolution (e.g.\ resolved by $\gtrsim10$ channels). Integrated SNR of $\lesssim30$ have been shown to cause the measured asymmetry of spectra to be larger than the intrinsic asymmetry with the effect decreasing with increasing SNR \citep{Watts2020}. We tabulate the number of HIPASS galaxies in subsamples after applying SNR cuts of 10 \citep[used in the literature, e.g.][]{Espada2011,Bok2019}, 20, 30 and 40 for the full velocity range and four velocity sub-intervals in Table~\ref{table:snr_cuts}. For our analysis we limit our sample to HIPASS galaxies with $\mathrm{SNR}>30$. Although this decreases the sample size to 1167 galaxies, it ensures that our results are not too strongly biased by the effect of SNR on the measured $A_{\mathrm{flux}}$.

\section{DISCUSSION}
\label{sec:discussion}

\subsection{Global HI Deficiency}
\label{s-sec:global_def}

\begin{figure*}
    \centering
    \includegraphics[width=17.5cm]{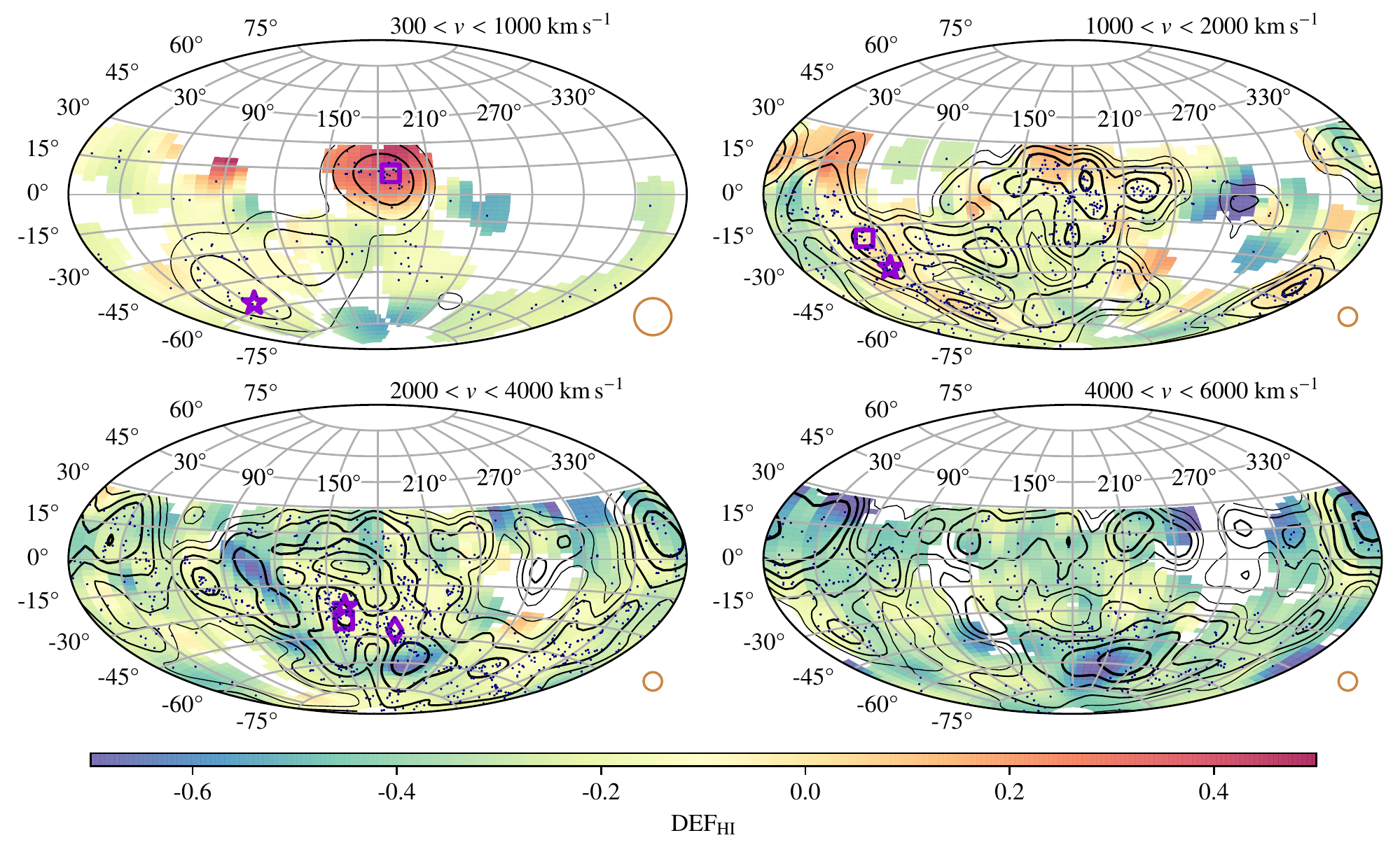}
		\caption{Gaussian weighted mean H\,\textsc{i} deficiency, $\mathrm{DEF}_{\mathrm{HI}}$, sky distribution plots. The Gaussian weighted mean $\mathrm{DEF}_{\mathrm{HI}}$ increases from low (blue) to high (red) values, with white indicating there is no data. The contour levels indicate the underlying galaxy density distribution from HyperLEDA of $\log(N/\mathrm{sr})=1.9$, 2.2, 2.5, 2.8, 3.1, 3.4, 3.7 for increasing line thickness. The points indicate the location of HIPASS galaxies. The orange circles in the lower right corner of each panel indicates the size of the Gaussian ($\sigma$) kernel used to smooth each map as it appears at the equator. The four panels show the velocity ranges: $V_1=300$--1\,000\,km\,s$^{-1}$, $V_2=1\,000$--2\,000\,km\,s$^{-1}$, $V_3=2\,000$--4\,000\,km\,s$^{-1}$ and $V_4=4\,000$--6\,000\,km\,s$^{-1}$ (top left, top right, bottom left and bottom right, respectively). The purple symbols indicate the positions of the Virgo cluster and Dorado group (square and star in top left panel, respectively), the Eridanus group and Fornax cluster (square and star in top right panel, respectively) and the Antlia, Hydra and Centaurus cluster regions (square, star and diamond in bottom left panel, respectively).}
	\label{fig:def_sky}
\end{figure*}

We produce the global sky distribution plots in Figure~\ref{fig:def_sky} using a different method to the one used by \cite{Denes2014}. We bin the environment density and H\,\textsc{i} galaxy parameters by convolving the sky with a 2-dimensional Gaussian and plot the mean H\,\textsc{i} parameter weighted by the Gaussian in each bin on the sky. \cite{Denes2014} directly binned the sky into spatial bins and plotted the weighted 2d histogram (i.e.\ the sum) of galaxies contained within each bin. The \cite{Denes2014} results using the weighted 2d histogram statistic are sensitive to highly H\,\textsc{i} rich and poor galaxies while we are interested in measuring the average H\,\textsc{i} deficiency and comparing with the average H\,\textsc{i} asymmetries (Section~\ref{s-sec:global_asym}). In this section we discuss similarities and differences between our results and those of \cite{Denes2014} using these two statistics. We create sky distribution plots for galaxies in the same four velocity ranges used in \cite{Denes2014}: $V_1=300$--1\,000\,km\,s$^{-1}$, $V_2=1\,000$--2\,000\,km\,s$^{-1}$, $V_3=2\,000$--4\,000\,km\,s$^{-1}$ and $V_4=4\,000$--6\,000\,km\,s$^{-1}$ (top left, top right, bottom left and bottom right panels, respectively, of Figure~\ref{fig:def_sky}). Throughout the rest of this paper we refer to the four velocity ranges as $V_1$, $V_2$, $V_3$ and $V_4$.

To create the sky distribution plot, we divide the sky into 50 bins in RA and Dec with the value in each bin corresponding to the Gaussian weighted mean H\,\textsc{i} deficiency at that sky position using a Gaussian with $\sigma=0.2$\,rad for $V_1$ and $\sigma=0.1$\,rad for $V_2$, $V_3$ and $V_4$. These smoothing kernels correspond to smoothing on physical scales of $\sim2$, $\sim2$, $\sim4.5$ and $\sim7.5$\,Mpc at the mean velocities of 650, 1\,500, 3\,000 and 5\,000\,km\,s$^{-1}$ in each velocity bin, respectively. Hence, while the $V_1$ and $V_2$ maps are smoothed to the same physical scale, the $V_3$ and $V_4$ maps are smoothed to roughly double and four times the physical scale of $V_1$ and $V_2$. We do not smooth $V_3$ and $V_4$ using a smaller kernel, as using a smaller smoothing kernel would result in unacceptably large statistical uncertainties. However, this does not affect our analysis as we do not draw conclusions from comparing between different velocity bins due to the different limiting magnitudes of the underlying HyperLEDA sample (see Section~\ref{s-sec:hyperleda}). We also investigate the effect that the smoothing scale has on our results for $\sigma=0.05$--0.30\,rad and find apart from the noisiness of the $\sigma=0.05$\,rad maps that the maps are generally insensitive to the choice of smoothing kernel (which is also the case for H\,\textsc{i} asymmetry sky distributions, Section~\ref{s-sec:global_asym}).

Similarly, we define the galaxy projected surface density from HyperLEDA at each sky position as the Gaussian weighted sum using the same Gaussian divided by the solid angle of a top hat with a radius equal to the Gaussian $\sigma$ ($\Omega=2\pi[1-\cos(\sigma)]$). This method has an advantage over standard binning, in which the value of each bin is derived in isolation, as all galaxies in our HIPASS and HyperLEDA subsamples contribute to each sky position, with more distant galaxies contributing less than nearby galaxies. We plot the sky distributions with weighted mean $\mathrm{DEF}_{\mathrm{HI}}$ in Figure~\ref{fig:def_sky} as shading (blue indicates more H\,\textsc{i} rich and red indicates more H\,\textsc{i} deficient, while white regions do not contain any data) with contours indicating the Gaussian weighted HyperLEDA projected surface density for the four velocity ranges ($V_1$, $V_2$, $V_3$, $V_4$) used in \cite{Denes2014}.

Qualitatively we find broad agreement between our sky distributions and those of \cite{Denes2014}, where discrepancies between this work and \cite{Denes2014} are due to differences in the binning method and the plotted statistics (weighted mean vs sum). As expected in the velocity range $V_1$, we find galaxies to be H\,\textsc{i} deficient in the region of the Virgo cluster ($\alpha,\delta=187.70^{\circ}, 12.34^{\circ}$, purple square), which is known to contain H\,\textsc{i} deficient galaxies \citep[e.g.][]{Chung2009}. We do not recover the summed H\,\textsc{i} deficiency for the Dorado group ($\alpha,\delta=64.27^{\circ}, -56.13^{\circ}$, purple star), however our result for the mean H\,\textsc{i} deficiency of $-0.2\lesssim\mathrm{DEF}_{\mathrm{HI}}\lesssim0.0$ (i.e.\ H\,\textsc{i} normal) is in agreement with the lack of a global H\,\textsc{i} deficiency in Dorado as determined by \cite{Kilborn2005}. We note that the H\,\textsc{i} deficiency signature that \cite{Denes2014} find is dominated by two galaxies, NGC\,1543 and NGC\,1546 ($\mathrm{DEF}_{\mathrm{HI}}=0.92$ and 0.72, respectively), and has a summed $\mathrm{DEF}_{\mathrm{HI}}$ of 1.9 (mean $\mathrm{DEF}_{\mathrm{HI}}$ is 0.24) including these two galaxies. If NGC\,1543 and NGC\,1546 are excluded then the summed $\mathrm{DEF}_{\mathrm{HI}}$ is 0.26, which is similar to the mean $\mathrm{DEF}_{\mathrm{HI}}$ including the galaxies. This illustrates the point that the results of \cite{Denes2014} are sensitive to outlier galaxies which are extremely H\,\textsc{i} rich or poor, while our results are sensitive to the average H\,\textsc{i} deficiency.

For $V_2$, the higher $\mathrm{DEF}_{\mathrm{HI}}$ values ($\mathrm{DEF}_{\mathrm{HI}} \gtrsim -0.1$) generally trace the regions with positive summed $\mathrm{DEF}_{\mathrm{HI}}$ values. The region around Virgo is the only location we find to be potentially H\,\textsc{i} deficient with $\mathrm{DEF}_{\mathrm{HI}}>0$, while the other regions \cite{Denes2014} highlighted (e.g.\ the Eridanus group, $\alpha,\delta=52.06^{\circ}, -20.74^{\circ}$, purple square, and the Fornax cluster, $\alpha,\delta=54.62^{\circ}, -35.45^{\circ}$, purple star), have normal H\,\textsc{i} content with mean deficiencies of $\mathrm{DEF}_{\mathrm{HI}}\sim0$. The lower mean H\,\textsc{i} deficiency we measure in Eridanus and Fornax compared to targeted surveys by \cite{Omar2005} and \cite{Schroder2001}, respectively, which found them to be H\,\textsc{i} deficient on average, is likely due to HIPASS, which is most sensitive to H\,\textsc{i} rich galaxies, not detecting the most H\,\textsc{i} deficient galaxies. 

HIPASS is even less sensitive to H\,\textsc{i} deficient galaxies in the higher velocity ranges, which likely contributes to our negative measured mean H\,\textsc{i} deficiency ($-0.20\lesssim\mathrm{DEF}_{\mathrm{HI}}\lesssim-0.15$) for the Antlia ($\alpha,\delta=157.51^{\circ}, -35.32^{\circ}$, purple square), Hydra ($\alpha,\delta=159.17^{\circ}, -27.52^{\circ}$, purple star) and Centaurus ($\alpha,\delta=192.20^{\circ}, -41.31^{\circ}$, purple diamond) cluster regions, indicating that they are on average H\,\textsc{i} normal, compared to positive summed values from \cite{Denes2014} at $V_3$. We have agreement with \cite{Denes2014} for the highest velocities ($V_4$), with all HIPASS detections being H\,\textsc{i} rich, which as \cite{Denes2014} point out, is expected for a blind H\,\textsc{i} survey.

\begin{figure*}
	\centering
	\includegraphics[width=14cm]{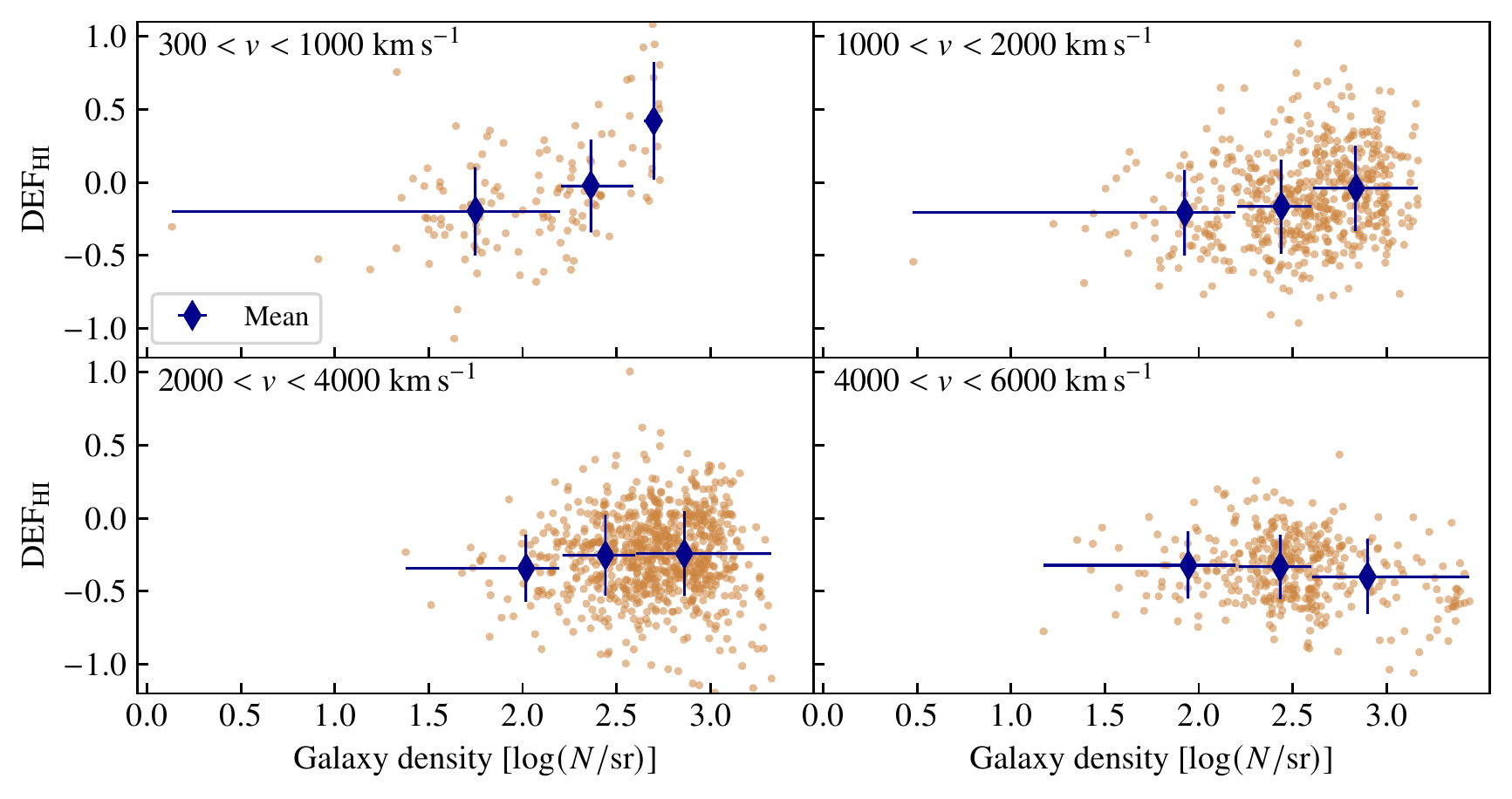}
		\caption{HIPASS galaxy H\,\textsc{i} deficiency, $\mathrm{DEF}_{\mathrm{HI}}$, vs projected surface density from HyperLEDA in the four velocity subsamples: $V_1$, $V_2$, $V_3$ and $V_4$ (top left, top right, bottom left and bottom right, respectively). The blue diamonds indicate the mean density and H\,\textsc{i} deficiency in each bin with vertical error bars showing the standard deviation (smaller than the mean symbols) and horizontal error bars showing the density range of galaxies in the bin.}
	\label{fig:def_density}
\end{figure*}

We also collapse the sky plots into 1d plots of $\mathrm{DEF}_{\mathrm{HI}}$ vs galaxy density in Figure~\ref{fig:def_density}. The density is calculated as in the sky plot, but now with the Gaussian centred on each HIPASS galaxy individually instead of the centre of each sky bin. In the two low velocity bins ($V_1$ and $V_2$, upper panels of Figure~\ref{fig:def_density}), we find clear trends of increasing mean H\,\textsc{i} deficiency with increasing density. This is most clear for $V_1$, where the mean for the highest density bin is $>0.3$, indicating that these galaxies contain on average less than half the expected H\,\textsc{i} and corresponds to the Virgo cluster region of the sky plot. For $V_2$, the mean increases across the three bins, but is never $>0.3$. There are however a number of individual galaxies in the two higher density bins which have $\mathrm{DEF}_{\mathrm{HI}}>0.3$. The two highest velocity ranges show no variation in the binned mean $\mathrm{DEF}_{\mathrm{HI}}$, which is $<0.0$ as expected as at these higher velocities HIPASS is most sensitive to H\,\textsc{i} rich and H\,\textsc{i} normal galaxies. The mean and standard deviation of $\mathrm{DEF}_{\mathrm{HI}}$ from Figure~\ref{fig:def_density} are tabulated in Table~\ref{table:hidef_mean}. We also find that the correlation, or lack thereof, between H\,\textsc{i} deficiency and environment density in each velocity range is insensitive to the size of the smoothing kernel ($\sigma=0.05$--0.30\,rad). This also holds true for the H\,\textsc{i} asymmetry analysis below (Section~\ref{s-sec:global_asym}).

\begin{table}
	\centering
    \caption{The mean and standard deviation of the H\,\textsc{i} deficiency, $\mathrm{DEF}_{\mathrm{HI}}$, for the three density bins ($\log(N/\mathrm{sr})<2.2$, 2.2--2.6, $>2.6$) in Figure~\ref{fig:def_density} and four the velocity subsamples.}
	\label{table:hidef_mean}
	\begin{tabular}{lccr}
		\hline
		Velocity & \multicolumn{3}{c}{Density [$\log(N/\mathrm{sr})$]} \\
		{[km\,s$^{-1}$]} & $<2.2$ & 2.2--2.6 & $>2.6$ \\ \hline
		300--1\,000    & $-0.19\pm0.30$ & $-0.02\pm0.32$ & $0.42\pm0.40$ \\
		1\,000--2\,000 & $-0.20\pm0.29$ & $-0.17\pm0.32$ & $-0.04\pm0.29$ \\
		2\,000--4\,000 & $-0.34\pm0.23$ & $-0.25\pm0.28$ & $-0.24\pm0.29$ \\
		4\,000--6\,000 & $-0.32\pm0.23$ & $-0.34\pm0.22$ & $-0.40\pm0.26$ \\ \hline
	\end{tabular}
\end{table}

\subsection{Global HI Asymmetry}
\label{s-sec:global_asym}

Studies using both spectral and morphological H\,\textsc{i} asymmetries find the fraction of galaxies that are asymmetric is larger in groups and clusters compared with isolated, field galaxies \citep[e.g.][]{Angiras2006,Angiras2007,Espada2011,Scott2018,Bok2019}. Using $\sim140$ galaxies from the LVHIS, VIVA and HALOGAS surveys, \cite{Reynolds2020} find signs of trends of increasing asymmetries with environment density. With HIPASS we can investigate the distribution of the H\,\textsc{i} flux asymmetry ratio, $A_{\mathrm{flux}}$, across the sky and look for the tentative trend in $A_{\mathrm{flux}}$ with environment that was reported by \cite{Reynolds2020} using a larger and more statistically robust galaxy sample.

\begin{figure*}
    \centering
    \includegraphics[width=17.5cm]{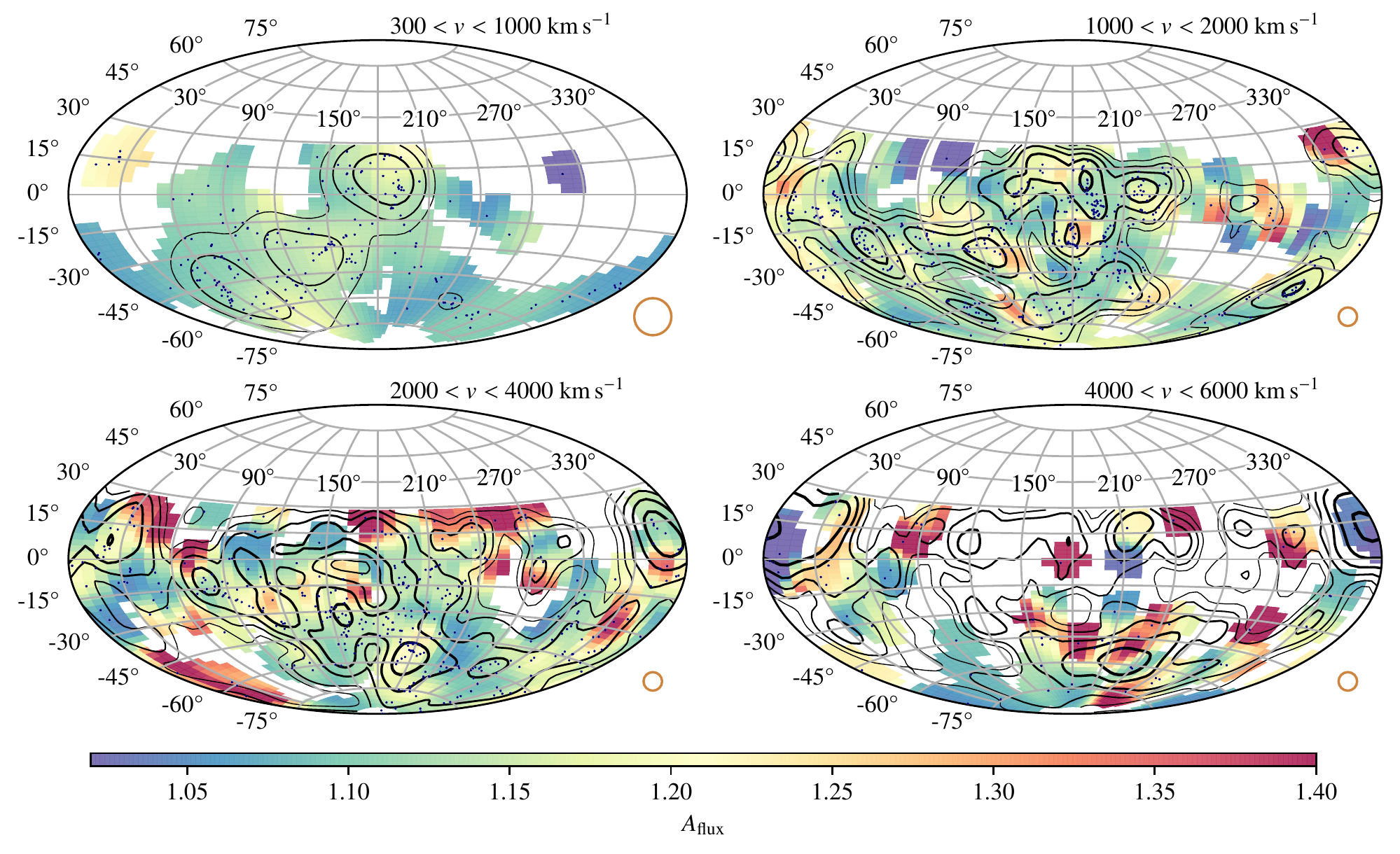}
	    \caption{Similar plot to Figure~\ref{fig:def_sky} but for the Gaussian weighted mean flux ratio asymmetry, $A_{\mathrm{flux}}$. The contour levels indicate the underlying galaxy density distribution from HyperLEDA of $\log(N/\mathrm{sr})=1.9$, 2.2, 2.5, 2.8, 3.1, 3.4, 3.7 for increasing line thickness. Note that the HIPASS galaxy subsamples do not correspond to those in Figure~\ref{fig:def_sky} due to the different selection criteria (e.g.\ $\mathrm{SNR}>30$ for $A_{\mathrm{flux}}$ and selection criteria from \citeauthor{Denes2014} \citeyear{Denes2014} for $\mathrm{DEF}_{\mathrm{HI}}$, Section~\ref{s-sec:hi_def}).}
	\label{fig:flux_sky}
\end{figure*}

We plot the sky distribution for weighted mean flux asymmetry ratio, $A_{\mathrm{flux}}$, in Figure~\ref{fig:flux_sky} following the same method described for the global H\,\textsc{i} deficiency sky distribution in Section~\ref{s-sec:global_def} (mean asymmetries increasing from blue to red shading). In the two low-velocity ranges ($V_1$ and $V_2$), the yellow and orange shading, which indicates higher mean asymmetries, tend to be coincident with the intermediate density HyperLEDA density contours. H\,\textsc{i} asymmetries do not appear to trace the density contours in the two highest velocity ranges.

\begin{figure*}
    \centering
    \includegraphics[width=14cm]{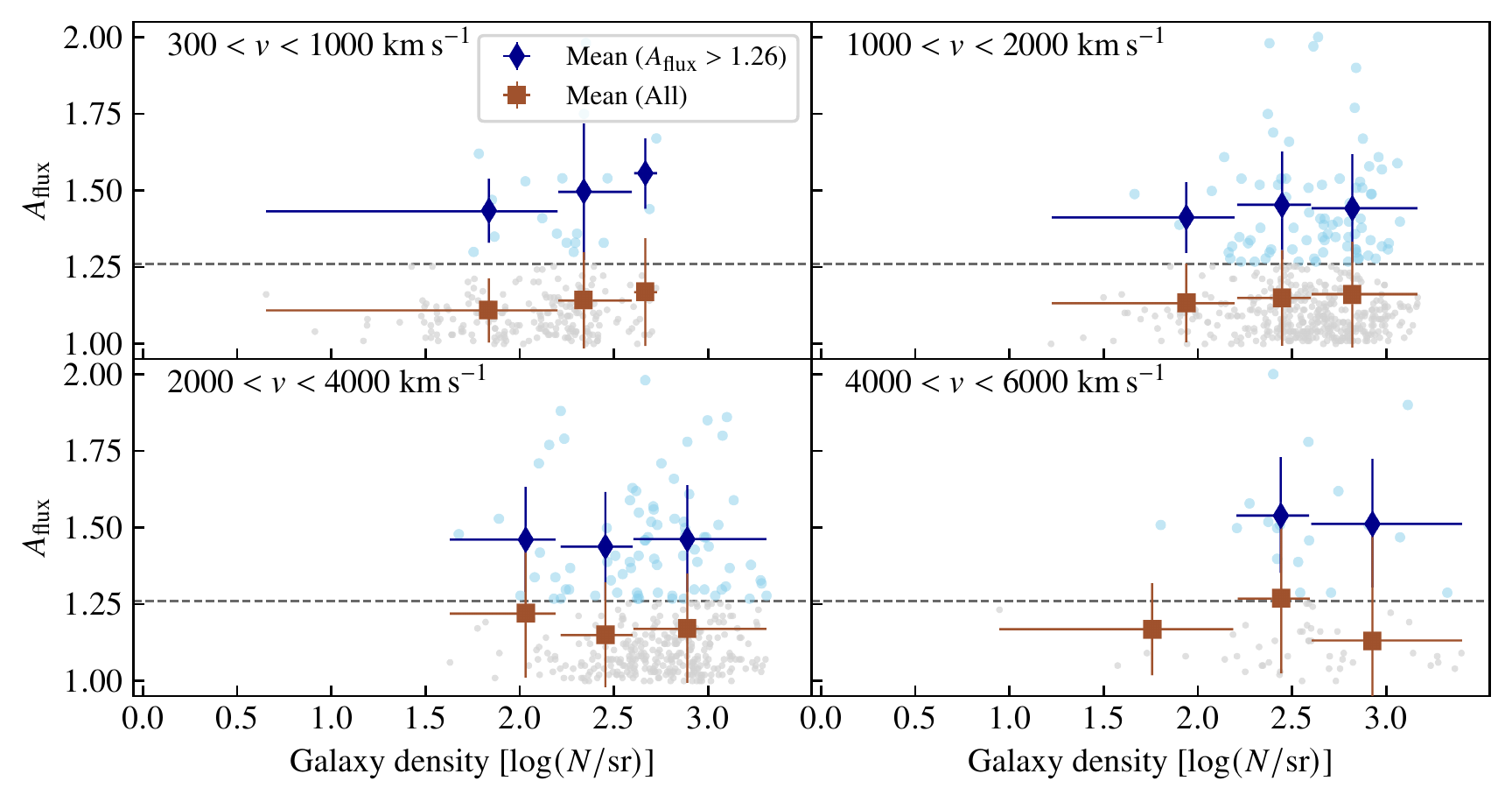}
	    \caption{HIPASS flux ratio asymmetries, $A_{\mathrm{flux}}$, vs galaxy density. The grey and blue points are the values for individual HIPASS galaxies with $A_{\mathrm{flux}}<1.26$ and $>1.26$, respectively (i.e.\ cut at the \citeauthor{Espada2011}, \citeyear{Espada2011} $2\sigma$ level, indicated by horizontal dashed line). The orange squares and blue diamonds indicate the mean density and $A_{\mathrm{flux}}$ for all galaxies and galaxies with $A_{\mathrm{flux}}>1.26$, respectively, in each bin with vertical error bars showing the standard deviation and horizontal error bars showing the density range of galaxies in the bin.}
	\label{fig:flux_density_2sig}
\end{figure*}

Similarly to $\mathrm{DEF}_{\mathrm{HI}}$, we also plot the individual HIPASS $A_{\mathrm{flux}}$ values against projected density computed using the HyperLEDA galaxies in each velocity range and the same Gaussian centre on each HIPASS galaxy in Figure~\ref{fig:flux_density_2sig}. The mean asymmetry for all HIPASS galaxies does not vary with density as measured in three density bins of $\log(N/\mathrm{sr})<2.2$, 2.2--2.6, $>2.6$ (Figure~\ref{fig:flux_density_2sig} and Table~\ref{table:percentiles}). For asymmetry analyses, of interest are the values of the outliers from the half Gaussian distribution of \cite{Espada2011} with statistically significant $A_{\mathrm{flux}}$ (e.g.\ $>2\sigma$) compared to isolated galaxies. For this reason we indicate the HIPASS galaxies with $A_{\mathrm{flux}}>1.26$ (denoted by the dashed line) in blue and measure the mean and standard deviation for only those galaxies with $A_{\mathrm{flux}}>1.26$ in the same density bins. In the low velocity bin, $V_1$, there is a trend of increasing mean $A_{\mathrm{flux}}$ with density, although this is not statistically significant as the variation in $A_{\mathrm{flux}}$ is within the $1\sigma$ uncertainties. Similar to the mean for all galaxies, there is no variation in the mean asymmetry for galaxies with $A_{\mathrm{flux}}>1.26$ (Figure~\ref{fig:flux_density_2sig} and Table~\ref{table:percentiles}). We note there is only one galaxy in the low density bin of $V_4$, hence there is no mean in this bin. 

\begin{table}
	\centering
    \caption{The fraction and mean and standard deviation of the flux ratio asymmetry, $A_{\mathrm{flux}}$, for galaxies with $A_{\mathrm{flux}}>1.26$ and mean and standard deviation for all galaxies in the three density bins in Figure~\ref{fig:flux_density_2sig} and the four velocity subsamples.}
	\label{table:percentiles}
	\begin{tabular}{lccr}
		\hline
		Velocity & \multicolumn{3}{c}{Density [$\log(N/\mathrm{sr})$]}  \\
		{[km\,s$^{-1}$]} & $<2.2$ & 2.2--2.6 & $>2.6$ \\ \hline
		\multicolumn{4}{c}{\bf{Fraction} ($A_{\mathrm{flux}}>1.26$)} \\
		300--1\,000    & $6\%$ (7/118) & $11\%$ (9/83) & $14\%$ (2/14) \\
		1\,000--2\,000 & $12\%$ (7/56) & $15\%$ (25/172) & $20\%$ (48/238) \\
		2\,000--4\,000 & $35\%$ (9/26) & $19\%$ (17/88) & $20\%$ (52/254) \\
		4\,000--6\,000 & $14\%$ (1/7) & $37\%$ (11/30) & $21\%$ (6/28) \\
		\multicolumn{4}{c}{\bf{Mean} ($A_{\mathrm{flux}}>1.26$)} \\
		300--1\,000 & $1.43\pm0.10$ & $1.50\pm0.22$ & $1.56\pm0.12$ \\
		1\,000--2\,000 & $1.41\pm0.11$ & $1.45\pm0.17$ & $1.44\pm0.18$ \\
		2\,000--4\,000 & $1.46\pm0.18$ & $1.44\pm0.18$ & $1.46\pm0.17$ \\
		4\,000--6\,000 & --- & $1.54\pm0.19$ & $1.51\pm0.21$ \\
		\multicolumn{4}{c}{\bf{Mean (All)}} \\
		300--1\,000 & $1.11\pm0.11$ & $1.14\pm0.16$ & $1.17\pm0.18$ \\
		1\,000--2\,000 & $1.13\pm0.13$ & $1.15\pm0.16$ & $1.16\pm0.17$ \\
		2\,000--4\,000 & $1.22\pm0.21$ & $1.15\pm0.17$ & $1.17\pm0.18$ \\
		4\,000--6\,000 & $1.17\pm0.15$ & $1.27\pm0.24$ & $1.13\pm0.41$ \\ \hline
	\end{tabular}
\end{table}

Previous studies have found asymmetric fractions of $\sim2$--$18\%$ in samples of isolated systems and $\sim16$--$27\%$ in pairs, groups and clusters for $A_{\mathrm{flux}}>1.26$ \citep{Haynes1998,Matthews1998,Espada2011,Scott2018,Bok2019}. We also calculate the fraction of galaxies with $A_{\mathrm{flux}}>1.26$ in the three density bins for the velocity subsamples (Table~\ref{table:percentiles}). We find a statistically significant increase in the asymmetric fraction ($A_{\mathrm{flux}}>1.26$) with density in the $V_2$ velocity bin from 12\%--20\% with uncertainties of $\pm3$--4\% (we assume a binomial distribution for calculating the uncertainties), which indicates that there is a correlation between asymmetric fraction and environment density in agreement with previous studies. The observed increase in the asymmetric fraction in $V_1$ (6\%--14\%) is within the uncertainties ($\pm2$--9\%) and is not statistically significant. There is no clear trend in the higher velocity bins between asymmetric fraction and environment density and the variation is within the uncertainties ($\pm3$--13\%). However, we note that we have limited our sample to galaxies with integrated $\mathrm{SNR}>30$ compared with the cut of $\mathrm{SNR}>10$ used in previous studies, which can explain the lower asymmetric fractions that we measure (see discussion of the effect of the SNR on the measured $A_{\mathrm{flux}}$ in Section~\ref{s-sec:asyms}). We also investigate varying the width of the velocity bins (e.g.\ $\Delta v=500$\,km\,s$^{-1}$ and $\Delta v=1\,000$\,km\,s$^{-1}$ starting from $v=500$\,km\,s$^{-1}$), however neither the mean nor the asymmetric fractions change within the uncertainties.

There may be a number of reasons why we do not observe an increase in the mean asymmetry in the intermediate density bin of Figure~\ref{fig:flux_density_2sig} that we observe in the sky plot, although these results are in agreement with the results of \cite{Reynolds2020}, who found the variation in the mean $A_{\mathrm{flux}}$ with environment density to be smaller than the sample $1\sigma$ standard deviation. Figure~\ref{fig:flux_sky} shows the local mean asymmetry, which varies with both low and high mean asymmetries coincident with the same density contours. Thus, when we average over the entire sky in Figure~\ref{fig:flux_density_2sig} the global mean will be lower than the higher local means.

A potentially greater effect in suppressing asymmetry variations is the method used to estimate the density around each HIPASS galaxy. Here we use 2-dimensional projected distances, which are less accurate than 3-dimensional distances for estimating densities \citep[e.g.][]{Reynolds2020}. We estimate densities using HyperLEDA galaxies in discrete velocity bins matched to the four HIPASS velocity ranges for consistency with the densities indicated in the sky plot \citep[density contours comparable to][]{Denes2014} and the 1d plot (Figures~\ref{fig:flux_sky} and \ref{fig:flux_density_2sig}, respectively). This also mitigates the effect of the variation in the HyperLEDA absolute magnitude limits (Table~\ref{table:absolute_magnitudes}). As density estimates used a window width between 700 and 2\,000\,km\,s$^{-1}$, this likely weakens any true correlation between asymmetry and density. We see this for the asymmetric fraction, in which we observe a positive correlation between asymmetric fraction and density in the $V_2$ velocity bin (spanning 1\,000\,km\,s$^{-1}$), but no correlation in $V_3$ or $V_4$ (spanning 2\,000\,km\,s$^{-1}$).

\begin{figure}
    \centering
    \includegraphics[width=\columnwidth]{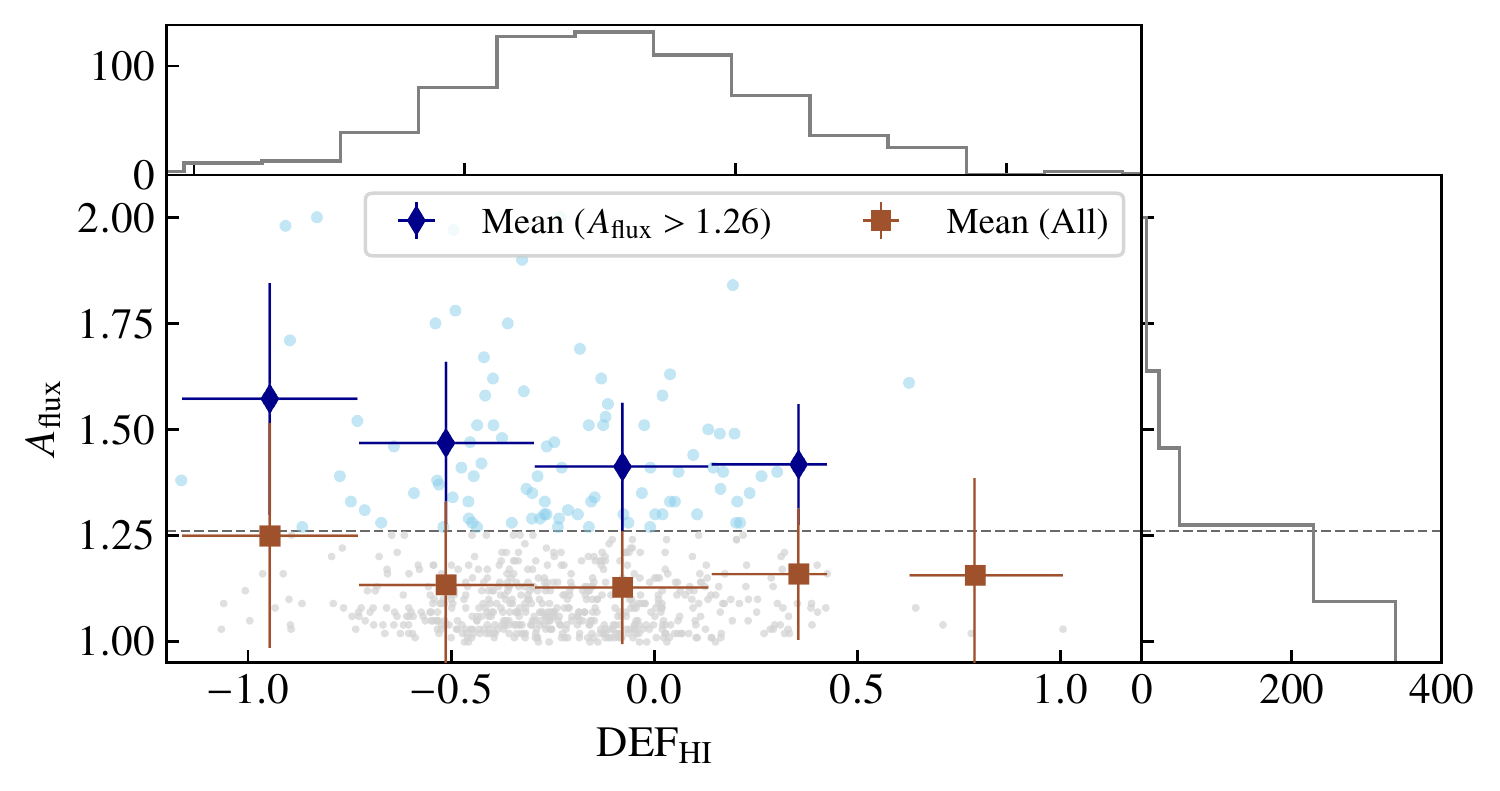}
	    \caption{Similar to Figure~\ref{fig:flux_density_2sig}, but for the flux asymmetry ratio, $A_{\mathrm{flux}}$, vs H\,\textsc{i} deficiency, $\mathrm{DEF}_{\mathrm{HI}}$ for all HIPASS galaxies within 300--6\,000\,km\,s$^{-1}$, with the histograms showing the distributions of $A_{\mathrm{flux}}$ and $\mathrm{DEF}_{\mathrm{HI}}$.}
	\label{fig:flux_def_2sigma}
\end{figure}

There may also be an underlying dependence on other galaxy properties, such as H\,\textsc{i} mass or H\,\textsc{i} deficiency ($\mathrm{DEF}_{\mathrm{HI}}$), that is contributing to the scatter observed in $A_{\mathrm{flux}}$ across all densities. For galaxies in the Virgo cluster, \cite{Scott2018} found the fraction (42\%) of low H\,\textsc{i} mass ($\log(M_{\mathrm{HI}}/\mathrm{M}_{\odot})\leq8.48$) galaxies with $A_{\mathrm{flux}}>1.39$ to be significantly higher than the fraction (16\%) of high H\,\textsc{i} mass galaxies ($\log(M_{\mathrm{HI}}/\mathrm{M}_{\odot})>8.48$). \cite{Scott2018} also identified a weak correlation between $A_{\mathrm{flux}}$ and $\mathrm{DEF}_{\mathrm{HI}}$ from 16 galaxies in Abell\,1367 with $2\sigma$ significance ($A_{\mathrm{flux}}>1.26$), although this correlation disappears if a $3\sigma$ cut ($A_{\mathrm{flux}}>1.39$) is applied. In Figure~\ref{fig:flux_def_2sigma}, we plot $A_{\mathrm{flux}}$ vs $\mathrm{DEF}_{\mathrm{HI}}$ for the HIPASS sample over the full velocity range 300--6\,000\,km\,s$^{-1}$ and measure the mean for five bins in $\mathrm{DEF}_{\mathrm{HI}}$ for the full sample and galaxies with $A_{\mathrm{flux}}>1.26$ and the fraction of galaxies in each bin with $A_{\mathrm{flux}}>1.26$ (Table~\ref{table:asym_def_mean}). We find no variation in the mean for all galaxies or those with $A_{\mathrm{flux}}>1.26$ with $\mathrm{DEF}_{\mathrm{HI}}$. We also find no variation in the asymmetric fraction with density. This implies that there is either no correlation between H\,\textsc{i} deficiency and asymmetry or, if a correlation does exist, it is subtle and we do not recover it due to the small sample size, which is biased towards H\,\textsc{i} rich galaxies (i.e.\ 55 galaxies of which $\sim3$ are classified as H\,\textsc{i} deficient with $\mathrm{DEF}_{\mathrm{HI}}>0.3$). We do not investigate other parameters (e.g.\ H\,\textsc{i} or stellar mass) due to HIPASS being a blind H\,\textsc{i} survey and our signal to noise ratio cut of 30. These characteristics bias our sample to galaxies with high H\,\textsc{i}, and subsequently stellar, mass (e.g.\ $\sim60\%$ of the sample have $9<\log(M_{\mathrm{HI}}/\mathrm{M}_{\odot})<10$ and $\sim85\%$ of the sample has $\log(M_{\mathrm{HI}}/\mathrm{M}_{\odot})>8.48$, the mass cut applied by \citeauthor{Scott2018} \citeyear{Scott2018}).

\begin{table}
	\centering
    \caption{The fraction and mean and standard deviation of the flux ratio asymmetry, $A_{\mathrm{flux}}$, for galaxies with $A_{\mathrm{flux}}>1.26$ and mean and standard deviation for all galaxies binned by H\,\textsc{i} deficiency, $\mathrm{DEF}_{\mathrm{HI}}$, from Figure~\ref{fig:flux_def_2sigma}.}
	\label{table:asym_def_mean}
	\begin{tabular}{lccr}
		\hline
		$\mathrm{DEF}_{\mathrm{HI}}$ & Fraction & $\langle A_{\mathrm{flux}} \rangle$ & $\langle A_{\mathrm{flux}} \rangle$ \\ 
		& $A_{\mathrm{flux}}>1.26$ & $>1.26$ & All \\ \hline
		$-0.95$ & $27\%$ (7/26)  & $1.57\pm0.27$ & $1.25\pm0.26$ \\
		$-0.51$ & $13\%$ (32/246) & $1.47\pm0.19$ & $1.13\pm0.20$ \\
		$-0.08$ & $12\%$ (38/314) & $1.41\pm0.15$ & $1.13\pm0.13$ \\
		$0.35$  & $20\%$ (12/61) & $1.42\pm0.14$ & $1.16\pm0.15$ \\ 
		$0.79$  & $20\%$ (1/5)    & ---           & $1.16\pm0.23$ \\ \hline
	\end{tabular}
\end{table}

\section{SUMMARY}
\label{sec:conclusion}

We have carried out a similar analysis to that of \cite{Denes2014} to investigate the variation of H\,\textsc{i} deficiency, $\mathrm{DEF}_{\mathrm{HI}}$, across the sky in the H\,\textsc{i} Parkes All-Sky Survey (HIPASS) and projected environment density estimated using HyperLEDA within four velocity ranges: $V_1=300$--1\,000\,km\,s$^{-1}$, $V_2=1\,000$--2\,000\,km\,s$^{-1}$, $V_3=2\,000$--4\,000\,km\,s$^{-1}$ and $V_4=4\,000$--6\,000\,km\,s$^{-1}$. \cite{Denes2014} binned the HIPASS deficiencies using a summed 2d histogram, which is sensitive to outliers with large H\,\textsc{i} deficiencies. We use the weighted mean deficiency, which down-weights all galaxies by $1/N$. The only region identified in \cite{Denes2014} as being H\,\textsc{i} deficient which we find to be on average H\,\textsc{i} deficient is the Virgo cluster in the 300--1\,000\,km\,s$^{-1}$ velocity range. We find all other regions identified as H\,\textsc{i} deficient in \cite{Denes2014} to be on average either H\,\textsc{i} normal or rich. However, even using the weighted mean H\,\textsc{i} deficiency, for the $V_1$ and $V_2$ ranges we find statistically significant trends of increasing H\,\textsc{i} deficiency with density, which supports the conclusion of \cite{Denes2014} that H\,\textsc{i} deficient galaxies tend to be in denser environments. These results illustrate the effect the statistic chosen for the analysis has on the interpretation.

We also apply the same method to investigate the global distribution of the mean flux asymmetry ratio, $A_{\mathrm{flux}}$. There appears to be some correspondence between regions with larger mean $A_{\mathrm{flux}}$ and intermediate densities in the global $A_{\mathrm{flux}}$ sky plot for galaxies with the velocity ranges: $V_1=300$--1\,000\,km\,s$^{-1}$, $V_2=1\,000$--2\,000\,km\,s$^{-1}$. While we do not find statistically significant correlations when the positional sky information is removed and the mean $A_{\mathrm{flux}}$ is measured with respect to density alone, we do observe a positive correlation between the fraction of galaxies with $A_{\mathrm{flux}}>1.26$ and density. The most likely causes washing out any underlying trend in the mean asymmetry are the averaging of galaxies across the sky, the influence of other galaxy properties (e.g.\ H\,\textsc{i} mass, stellar mass and H\,\textsc{i} deficiency), estimating the environment density using 2d projected distances and only using HyperLEDA galaxies within a fixed velocity range for all HIPASS galaxies in each velocity bin for estimating density. We find no correlation between $A_{\mathrm{flux}}$ and $\mathrm{DEF}_{\mathrm{HI}}$, which implies that asymmetries do not depend on H\,\textsc{i} deficiency or the dependence is too subtle to be recovered with HIPASS.

In the future it will be possible to carry out a similar analysis for morphological and kinematic asymmetries using the WALLABY survey, which will spatially resolve all galaxies detected in HIPASS \citep{Koribalski2020}. This will also produce integrated spectra with higher signal to noise ratios, which will increase the sample for analysing spectral asymmetries. WALLABY will also provide $\sim1$--2 orders of magnitude more galaxies across a range of environments and, while still a shallow survey, will detect more low mass and H\,\textsc{i} deficient galaxies than possible with HIPASS. The increased number of detections will provide a more statistically robust sample for measuring the global distribution of H\,\textsc{i} spectral asymmetries and investigate correlations between asymmetries and other galaxy properties, such as H\,\textsc{i} deficiency and H\,\textsc{i} and stellar mass. 

\section*{Acknowledgements}

We thank S. Blythe and B. Holwerda for providing helpful comments. This research was conducted by the Australian Research Council Centre of Excellence for All Sky Astrophysics in 3 Dimensions (ASTRO 3D), through project number CE170100013. The Parkes radio telescope is part of the Australia Telescope National Facility which is funded by the Commonwealth of Australia for operation as a National Facility managed by CSRIO. We acknowledge the usage of the HyperLEDA database (\url{http://leda.univ-lyon1.fr}).

\section*{Data Availability}

The HIPASS spectral line cubes are accessible on the CSIRO ASKAP Science Data Archive (CASDA) which can be accessed by following the DOI: \url{https://doi.org/10.25919/5c36de6d37141}.




\bibliographystyle{mnras}
\bibliography{master}

\begin{thebibliography}{}
\makeatletter
\relax
\def\mn@urlcharsother{\let\do\@makeother \do\$\do\&\do\#\do\^\do\_\do\%\do\~}
\def\mn@doi{\begingroup\mn@urlcharsother \@ifnextchar [ {\mn@doi@}
  {\mn@doi@[]}}
\def\mn@doi@[#1]#2{\def\@tempa{#1}\ifx\@tempa\@empty \href
  {http://dx.doi.org/#2} {doi:#2}\else \href {http://dx.doi.org/#2} {#1}\fi
  \endgroup}
\def\mn@eprint#1#2{\mn@eprint@#1:#2::\@nil}
\def\mn@eprint@arXiv#1{\href {http://arxiv.org/abs/#1} {{\tt arXiv:#1}}}
\def\mn@eprint@dblp#1{\href {http://dblp.uni-trier.de/rec/bibtex/#1.xml}
  {dblp:#1}}
\def\mn@eprint@#1:#2:#3:#4\@nil{\def\@tempa {#1}\def\@tempb {#2}\def\@tempc
  {#3}\ifx \@tempc \@empty \let \@tempc \@tempb \let \@tempb \@tempa \fi \ifx
  \@tempb \@empty \def\@tempb {arXiv}\fi \@ifundefined
  {mn@eprint@\@tempb}{\@tempb:\@tempc}{\expandafter \expandafter \csname
  mn@eprint@\@tempb\endcsname \expandafter{\@tempc}}}

\bibitem[\protect\citeauthoryear{Angiras, Jog, Omar  \& Dwarakanath}{Angiras
  et~al.}{2006}]{Angiras2006}
Angiras R.~A.,  Jog C.~J.,  Omar A.,   Dwarakanath K.~S.,  2006, \mn@doi
  [\mnras] {10.1111/j.1365-2966.2006.10418.x}, 369, 1849

\bibitem[\protect\citeauthoryear{Angiras, Jog, Dwarakanath  \&
  Verheijen}{Angiras et~al.}{2007}]{Angiras2007}
Angiras R.~A.,  Jog C.~J.,  Dwarakanath K.~S.,   Verheijen M.~A.,  2007,
  \mn@doi [\mnras] {10.1111/j.1365-2966.2007.11779.x}, 378, 276

\bibitem[\protect\citeauthoryear{{Barnes} et~al.,}{{Barnes}
  et~al.}{2001}]{Barnes2001}
{Barnes} D.~G.,  et~al., 2001, \mn@doi [\mnras]
  {10.1046/j.1365-8711.2001.04102.x}, \href
  {http://adsabs.harvard.edu/abs/2001MNRAS.322..486B} {322, 486}

\bibitem[\protect\citeauthoryear{{Bok}, {Blyth}, {Gilbank}  \& {Elson}}{{Bok}
  et~al.}{2019}]{Bok2019}
{Bok} J.,  {Blyth} S.~L.,  {Gilbank} D.~G.,   {Elson} E.~C.,  2019, \mn@doi
  [\mnras] {10.1093/mnras/sty3448}, \href
  {https://ui.adsabs.harvard.edu/abs/2019MNRAS.484..582B} {484, 582}

\bibitem[\protect\citeauthoryear{Boselli \& Gavazzi}{Boselli \&
  Gavazzi}{2009}]{Boselli2009}
Boselli A.,  Gavazzi G.,  2009, \mn@doi [aap] {10.1051/0004-6361/200912658},
  508, 201

\bibitem[\protect\citeauthoryear{Bournaud, Combes, Jog  \& Puerari}{Bournaud
  et~al.}{2005}]{Bournaud2005b}
Bournaud F.,  Combes F.,  Jog C.~J.,   Puerari I.,  2005, \mn@doi [Astronomy
  {\&} Astrophysics] {10.1051/0004-6361:20052631}, 438, 507

\bibitem[\protect\citeauthoryear{{Bravo-Alfaro}, {Cayatte}, {van Gorkom}  \&
  {Balkowski}}{{Bravo-Alfaro} et~al.}{2000}]{BravoAlfaro2000}
{Bravo-Alfaro} H.,  {Cayatte} V.,  {van Gorkom} J.~H.,   {Balkowski} C.,  2000,
  \mn@doi [\aj] {10.1086/301194}, \href
  {https://ui.adsabs.harvard.edu/abs/2000AJ....119..580B} {119, 580}

\bibitem[\protect\citeauthoryear{{Chamaraux}, {Balkowski}  \&
  {Gerard}}{{Chamaraux} et~al.}{1980}]{Chamaraux1980}
{Chamaraux} P.,  {Balkowski} C.,   {Gerard} E.,  1980, \aap, \href
  {https://ui.adsabs.harvard.edu/abs/1980A&A....83...38C} {83, 38}

\bibitem[\protect\citeauthoryear{{Chung}, {van Gorkom}, {Kenney}, {Crowl}  \&
  {Vollmer}}{{Chung} et~al.}{2009}]{Chung2009}
{Chung} A.,  {van Gorkom} J.~H.,  {Kenney} J.~D.~P.,  {Crowl} H.,   {Vollmer}
  B.,  2009, \mn@doi [\aj] {10.1088/0004-6256/138/6/1741}, \href
  {http://adsabs.harvard.edu/abs/2009AJ....138.1741C} {138, 1741}

\bibitem[\protect\citeauthoryear{Cortese, Catinella, Boissier, Boselli  \&
  Heinis}{Cortese et~al.}{2011}]{Cortese2011}
Cortese L.,  Catinella B.,  Boissier S.,  Boselli A.,   Heinis S.,  2011,
  \mn@doi [\mnras] {10.1111/j.1365-2966.2011.18822.x}, 415, 1797

\bibitem[\protect\citeauthoryear{{D{\'e}nes}, {Kilborn}  \&
  {Koribalski}}{{D{\'e}nes} et~al.}{2014}]{Denes2014}
{D{\'e}nes} H.,  {Kilborn} V.~A.,   {Koribalski} B.~S.,  2014, \mn@doi [\mnras]
  {10.1093/mnras/stu1337}, \href
  {http://adsabs.harvard.edu/abs/2014MNRAS.444..667D} {444, 667}

\bibitem[\protect\citeauthoryear{{D{\'e}nes}, {Kilborn}, {Koribalski}  \&
  {Wong}}{{D{\'e}nes} et~al.}{2016}]{Denes2016}
{D{\'e}nes} H.,  {Kilborn} V.~A.,  {Koribalski} B.~S.,   {Wong} O.~I.,  2016,
  \mn@doi [\mnras] {10.1093/mnras/stv2391}, \href
  {http://adsabs.harvard.edu/abs/2016MNRAS.455.1294D} {455, 1294}

\bibitem[\protect\citeauthoryear{{Doyle} et~al.,}{{Doyle}
  et~al.}{2005}]{Doyle2005}
{Doyle} M.~T.,  et~al., 2005, \mn@doi [\mnras]
  {10.1111/j.1365-2966.2005.09159.x}, \href
  {https://ui.adsabs.harvard.edu/abs/2005MNRAS.361...34D} {361, 34}

\bibitem[\protect\citeauthoryear{{Dressler}}{{Dressler}}{1980}]{Dressler1980}
{Dressler} A.,  1980, \mn@doi [\apj] {10.1086/157753}, \href
  {http://adsabs.harvard.edu/abs/1980ApJ...236..351D} {236, 351}

\bibitem[\protect\citeauthoryear{{English}, {Koribalski}, {Bland-Hawthorn},
  {Freeman}  \& {McCain}}{{English} et~al.}{2010}]{English2010}
{English} J.,  {Koribalski} B.,  {Bland-Hawthorn} J.,  {Freeman} K.~C.,
  {McCain} C.~F.,  2010, \mn@doi [\aj] {10.1088/0004-6256/139/1/102}, \href
  {http://adsabs.harvard.edu/abs/2010AJ....139..102E} {139, 102}

\bibitem[\protect\citeauthoryear{Espada, Verdes-Montenegro, Huchtmeier,
  Sulentic, Verley, Leon  \& Sabater}{Espada et~al.}{2011}]{Espada2011}
Espada D.,  Verdes-Montenegro L.,  Huchtmeier W.~K.,  Sulentic J.,  Verley S.,
  Leon S.,   Sabater J.,  2011, \mn@doi [Astronomy {\&} Astrophysics]
  {10.1051/0004-6361/201016117}, 532, A117

\bibitem[\protect\citeauthoryear{{Giovanelli} \& {Haynes}}{{Giovanelli} \&
  {Haynes}}{1985}]{Giovanelli1985}
{Giovanelli} R.,  {Haynes} M.~P.,  1985, \mn@doi [\apj] {10.1086/163170}, \href
  {http://adsabs.harvard.edu/abs/1985ApJ...292..404G} {292, 404}

\bibitem[\protect\citeauthoryear{{Giuricin}, {Marinoni}, {Ceriani}  \&
  {Pisani}}{{Giuricin} et~al.}{2000}]{Giuricin2000}
{Giuricin} G.,  {Marinoni} C.,  {Ceriani} L.,   {Pisani} A.,  2000, \mn@doi
  [\apj] {10.1086/317070}, \href
  {https://ui.adsabs.harvard.edu/abs/2000ApJ...543..178G} {543, 178}

\bibitem[\protect\citeauthoryear{{Gourgoulhon}, {Chamaraux}  \&
  {Fouque}}{{Gourgoulhon} et~al.}{1992}]{Gourgoulhon1992}
{Gourgoulhon} E.,  {Chamaraux} P.,   {Fouque} P.,  1992, \aap, \href
  {http://adsabs.harvard.edu/abs/1992A%26A...255...69G} {255, 69}

\bibitem[\protect\citeauthoryear{{Gunn} \& {Gott}}{{Gunn} \&
  {Gott}}{1972}]{Gunn1972}
{Gunn} J.~E.,  {Gott} III J.~R.,  1972, \mn@doi [\apj] {10.1086/151605}, \href
  {http://adsabs.harvard.edu/abs/1972ApJ...176....1G} {176, 1}

\bibitem[\protect\citeauthoryear{Haynes \& Giovanelli}{Haynes \&
  Giovanelli}{1984}]{Haynes1984}
Haynes M.~P.,  Giovanelli R.,  1984, The Astronomical Journal, 89

\bibitem[\protect\citeauthoryear{Haynes, Hogg, Maddalena, Roberts  \& van
  Zee}{Haynes et~al.}{1998}]{Haynes1998}
Haynes M.~P.,  Hogg D.~E.,  Maddalena R.~J.,  Roberts M.~S.,   van Zee L.,
  1998, \mn@doi [The Astronomical Journal] {10.1086/300166}, 115, 62

\bibitem[\protect\citeauthoryear{Hess \& Wilcots}{Hess \&
  Wilcots}{2013}]{Hess2013}
Hess K.~M.,  Wilcots E.~M.,  2013, \mn@doi [Astronomical Journal]
  {10.1088/0004-6256/146/5/124}, 146

\bibitem[\protect\citeauthoryear{{Johnston} et~al.,}{{Johnston}
  et~al.}{2007}]{Johnston2007}
{Johnston} S.,  et~al., 2007, \mn@doi [\pasa] {10.1071/AS07033}, \href
  {http://adsabs.harvard.edu/abs/2007PASA...24..174J} {24, 174}

\bibitem[\protect\citeauthoryear{{Jones}, {Haynes}, {Giovanelli}  \&
  {Papastergis}}{{Jones} et~al.}{2016}]{Jones2016}
{Jones} M.~G.,  {Haynes} M.~P.,  {Giovanelli} R.,   {Papastergis} E.,  2016,
  \mn@doi [\mnras] {10.1093/mnras/stv2394}, \href
  {http://adsabs.harvard.edu/abs/2016MNRAS.455.1574J} {455, 1574}

\bibitem[\protect\citeauthoryear{Kilborn, Koribalski, Forbes, Barnes  \&
  Musgrave}{Kilborn et~al.}{2005}]{Kilborn2005}
Kilborn V.~A.,  Koribalski B.~S.,  Forbes D.~A.,  Barnes D.~G.,   Musgrave
  R.~C.,  2005, \mn@doi [\mnras] {10.1111/j.1365-2966.2004.08450.x}, 356, 77

\bibitem[\protect\citeauthoryear{Kilborn, Forbes, Barnes, Koribalski, Brough
  \& Kern}{Kilborn et~al.}{2009}]{Kilborn2009}
Kilborn V.~A.,  Forbes D.~A.,  Barnes D.~G.,  Koribalski B.~S.,  Brough S.,
  Kern K.,  2009, \mn@doi [\mnras] {10.1111/j.1365-2966.2009.15587.x}, 400,
  1962

\bibitem[\protect\citeauthoryear{{Koribalski} \&
  {L{\'o}pez-S{\'a}nchez}}{{Koribalski} \&
  {L{\'o}pez-S{\'a}nchez}}{2009}]{Koribalski2009}
{Koribalski} B.~S.,  {L{\'o}pez-S{\'a}nchez} {\'A}.~R.,  2009, \mn@doi [\mnras]
  {10.1111/j.1365-2966.2009.15610.x}, \href
  {http://adsabs.harvard.edu/abs/2009MNRAS.400.1749K} {400, 1749}

\bibitem[\protect\citeauthoryear{{Koribalski} et~al.,}{{Koribalski}
  et~al.}{2020}]{Koribalski2020}
{Koribalski} B.~S.,  et~al., 2020, \mn@doi [\apss]
  {10.1007/s10509-020-03831-4}, \href
  {https://ui.adsabs.harvard.edu/abs/2020Ap&SS.365..118K} {365, 118}

\bibitem[\protect\citeauthoryear{{Makarov}, {Prugniel}, {Terekhova}, {Courtois}
   \& {Vauglin}}{{Makarov} et~al.}{2014}]{Makarov2014}
{Makarov} D.,  {Prugniel} P.,  {Terekhova} N.,  {Courtois} H.,   {Vauglin} I.,
  2014, \mn@doi [\aap] {10.1051/0004-6361/201423496}, \href
  {http://adsabs.harvard.edu/abs/2014A%26A...570A..13M} {570, A13}

\bibitem[\protect\citeauthoryear{Matthews, van Driel  \& {Gallagher
  III}}{Matthews et~al.}{1998}]{Matthews1998}
Matthews L.~D.,  van Driel W.,   {Gallagher III} J.~S.,  1998, \mn@doi [The
  Astronomical Journal] {10.1086/300492}, 116, 1169

\bibitem[\protect\citeauthoryear{{Meyer} et~al.,}{{Meyer}
  et~al.}{2004}]{Meyer2004}
{Meyer} M.~J.,  et~al., 2004, \mn@doi [\mnras]
  {10.1111/j.1365-2966.2004.07710.x}, \href
  {http://adsabs.harvard.edu/abs/2004MNRAS.350.1195M} {350, 1195}

\bibitem[\protect\citeauthoryear{{Moore}, {Katz}, {Lake}, {Dressler}  \&
  {Oemler}}{{Moore} et~al.}{1996}]{Moore1996}
{Moore} B.,  {Katz} N.,  {Lake} G.,  {Dressler} A.,   {Oemler} A.,  1996,
  \mn@doi [\nat] {10.1038/379613a0}, \href
  {http://adsabs.harvard.edu/abs/1996Natur.379..613M} {379, 613}

\bibitem[\protect\citeauthoryear{{Moore}, {Lake}  \& {Katz}}{{Moore}
  et~al.}{1998}]{Moore1998}
{Moore} B.,  {Lake} G.,   {Katz} N.,  1998, \mn@doi [\apj] {10.1086/305264},
  \href {http://adsabs.harvard.edu/abs/1998ApJ...495..139M} {495, 139}

\bibitem[\protect\citeauthoryear{{Moore}, {Lake}, {Quinn}  \& {Stadel}}{{Moore}
  et~al.}{1999}]{Moore1999}
{Moore} B.,  {Lake} G.,  {Quinn} T.,   {Stadel} J.,  1999, \mn@doi [\mnras]
  {10.1046/j.1365-8711.1999.02345.x}, \href
  {http://adsabs.harvard.edu/abs/1999MNRAS.304..465M} {304, 465}

\bibitem[\protect\citeauthoryear{{Muldrew} et~al.,}{{Muldrew}
  et~al.}{2012}]{Muldrew2012}
{Muldrew} S.~I.,  et~al., 2012, \mn@doi [\mnras]
  {10.1111/j.1365-2966.2011.19922.x}, \href
  {https://ui.adsabs.harvard.edu/abs/2012MNRAS.419.2670M} {419, 2670}

\bibitem[\protect\citeauthoryear{Odekon et~al.,}{Odekon
  et~al.}{2016}]{Odekon2016}
Odekon M.~C.,  et~al., 2016, \mn@doi [The Astrophysical Journal]
  {10.3847/0004-637X/824/2/110}, 824, 1

\bibitem[\protect\citeauthoryear{{Omar} \& {Dwarakanath}}{{Omar} \&
  {Dwarakanath}}{2005}]{Omar2005}
{Omar} A.,  {Dwarakanath} K.~S.,  2005, \mn@doi [Journal of Astrophysics and
  Astronomy] {10.1007/BF02702452}, \href
  {https://ui.adsabs.harvard.edu/abs/2005JApA...26...71O} {26, 71}

\bibitem[\protect\citeauthoryear{{Paturel}, {Theureau}, {Bottinelli},
  {Gouguenheim}, {Coudreau-Durand}, {Hallet}  \& {Petit}}{{Paturel}
  et~al.}{2003}]{Paturel2003}
{Paturel} G.,  {Theureau} G.,  {Bottinelli} L.,  {Gouguenheim} L.,
  {Coudreau-Durand} N.,  {Hallet} N.,   {Petit} C.,  2003, \mn@doi [\aap]
  {10.1051/0004-6361:20031412}, \href
  {https://ui.adsabs.harvard.edu/abs/2003A&A...412...57P} {412, 57}

\bibitem[\protect\citeauthoryear{{Planck Collaboration} et~al.,}{{Planck
  Collaboration} et~al.}{2016}]{Planck2016}
{Planck Collaboration} et~al., 2016, \mn@doi [\aap]
  {10.1051/0004-6361/201525830}, \href
  {http://adsabs.harvard.edu/abs/2016A%26A...594A..13P} {594, A13}

\bibitem[\protect\citeauthoryear{{Rasmussen}, {Ponman}  \&
  {Mulchaey}}{{Rasmussen} et~al.}{2006}]{Rasmussen2006}
{Rasmussen} J.,  {Ponman} T.~J.,   {Mulchaey} J.~S.,  2006, \mn@doi [\mnras]
  {10.1111/j.1365-2966.2006.10492.x}, \href
  {http://adsabs.harvard.edu/abs/2006MNRAS.370..453R} {370, 453}

\bibitem[\protect\citeauthoryear{{Rasmussen} et~al.,}{{Rasmussen}
  et~al.}{2012}]{Rasmussen2012}
{Rasmussen} J.,  et~al., 2012, \mn@doi [\apj] {10.1088/0004-637X/747/1/31},
  \href {http://adsabs.harvard.edu/abs/2012ApJ...747...31R} {747, 31}

\bibitem[\protect\citeauthoryear{{Reynolds}, {Westmeier}, {Staveley-Smith},
  {Chauhan}  \& {Lagos}}{{Reynolds} et~al.}{2020}]{Reynolds2020}
{Reynolds} T.~N.,  {Westmeier} T.,  {Staveley-Smith} L.,  {Chauhan} G.,
  {Lagos} C.~D.~P.,  2020, \mn@doi [\mnras] {10.1093/mnras/staa597}, \href
  {https://ui.adsabs.harvard.edu/abs/2020MNRAS.493.5089R} {493, 5089}

\bibitem[\protect\citeauthoryear{Richter \& Sancisi}{Richter \&
  Sancisi}{1994}]{Richter1994}
Richter O.-G.,  Sancisi R.,  1994, $\backslash$Aap, 290, L9

\bibitem[\protect\citeauthoryear{{Schr{\"o}der}, {Drinkwater}  \&
  {Richter}}{{Schr{\"o}der} et~al.}{2001}]{Schroder2001}
{Schr{\"o}der} A.,  {Drinkwater} M.~J.,   {Richter} O.~G.,  2001, \mn@doi
  [\aap] {10.1051/0004-6361:20010997}, \href
  {https://ui.adsabs.harvard.edu/abs/2001A&A...376...98S} {376, 98}

\bibitem[\protect\citeauthoryear{Scott, Brinks, Cortese, Bosell  \&
  Bravo-Alfaro}{Scott et~al.}{2018}]{Scott2018}
Scott T.~C.,  Brinks E.,  Cortese L.,  Bosell A.,   Bravo-Alfaro H.,  2018,
  \mn@doi [Monthly Notices of the Royal Astronomical Society]
  {10.1093/mnras/sty063}, pp 1--24

\bibitem[\protect\citeauthoryear{{Solanes}, {Manrique},
  {Garc{\'{\i}}a-G{\'o}mez}, {Gonz{\'a}lez-Casado}, {Giovanelli}  \&
  {Haynes}}{{Solanes} et~al.}{2001}]{Solanes2001}
{Solanes} J.~M.,  {Manrique} A.,  {Garc{\'{\i}}a-G{\'o}mez} C.,
  {Gonz{\'a}lez-Casado} G.,  {Giovanelli} R.,   {Haynes} M.~P.,  2001, \mn@doi
  [\apj] {10.1086/318672}, \href
  {http://adsabs.harvard.edu/abs/2001ApJ...548...97S} {548, 97}

\bibitem[\protect\citeauthoryear{{Tully}}{{Tully}}{1987}]{Tully1987}
{Tully} R.~B.,  1987, \mn@doi [\apj] {10.1086/165629}, \href
  {http://adsabs.harvard.edu/abs/1987ApJ...321..280T} {321, 280}

\bibitem[\protect\citeauthoryear{{Verdes-Montenegro}, {Yun}, {Williams},
  {Huchtmeier}, {Del Olmo}  \& {Perea}}{{Verdes-Montenegro}
  et~al.}{2001}]{Verdes-Montenegro2001}
{Verdes-Montenegro} L.,  {Yun} M.~S.,  {Williams} B.~A.,  {Huchtmeier} W.~K.,
  {Del Olmo} A.,   {Perea} J.,  2001, \mn@doi [\aap]
  {10.1051/0004-6361:20011127}, \href
  {http://adsabs.harvard.edu/abs/2001A%26A...377..812V} {377, 812}

\bibitem[\protect\citeauthoryear{{Verdes-Montenegro}, {Sulentic}, {Lisenfeld},
  {Leon}, {Espada}, {Garcia}, {Sabater}  \& {Verley}}{{Verdes-Montenegro}
  et~al.}{2005}]{VerdesMontenegro2005}
{Verdes-Montenegro} L.,  {Sulentic} J.,  {Lisenfeld} U.,  {Leon} S.,  {Espada}
  D.,  {Garcia} E.,  {Sabater} J.,   {Verley} S.,  2005, \mn@doi [\aap]
  {10.1051/0004-6361:20042280}, \href
  {https://ui.adsabs.harvard.edu/abs/2005A&A...436..443V} {436, 443}

\bibitem[\protect\citeauthoryear{Watts, Catinella, Cortese  \& Power}{Watts
  et~al.}{2020}]{Watts2020}
Watts A.,  Catinella B.,  Cortese L.,   Power C.,  2020, \mn@doi [\mnras]
  {10.1093/mnras/staa094}

\bibitem[\protect\citeauthoryear{Westmeier, Braun  \& Koribalski}{Westmeier
  et~al.}{2011}]{Westmeier2011}
Westmeier T.,  Braun R.,   Koribalski B.~S.,  2011, \mn@doi [\mnras]
  {doi:10.1111/j.1365-2966.2010.17596.x}, 410, 2217

\bibitem[\protect\citeauthoryear{{Wong} et~al.,}{{Wong}
  et~al.}{2006}]{Wong2006}
{Wong} O.~I.,  et~al., 2006, \mn@doi [\mnras]
  {10.1111/j.1365-2966.2006.10846.x}, \href
  {https://ui.adsabs.harvard.edu/abs/2006MNRAS.371.1855W} {371, 1855}

\bibitem[\protect\citeauthoryear{{Wong}, {Webster}, {Kilborn}, {Waugh}  \&
  {Staveley-Smith}}{{Wong} et~al.}{2009}]{Wong2009}
{Wong} O.~I.,  {Webster} R.~L.,  {Kilborn} V.~A.,  {Waugh} M.,
  {Staveley-Smith} L.,  2009, \mn@doi [\mnras]
  {10.1111/j.1365-2966.2009.15436.x}, \href
  {https://ui.adsabs.harvard.edu/abs/2009MNRAS.399.2264W} {399, 2264}

\bibitem[\protect\citeauthoryear{Yoon, Chung, Smith  \& Jaff{\'{e}}}{Yoon
  et~al.}{2017}]{Yoon2017}
Yoon H.,  Chung A.,  Smith R.,   Jaff{\'{e}} Y.~L.,  2017, \mn@doi [\aj]
  {10.3847/1538-4357/aa6579}, 838, 81

\bibitem[\protect\citeauthoryear{Zaritsky \& Rix}{Zaritsky \&
  Rix}{1997}]{Zaritsky1997}
Zaritsky D.,  Rix H.,  1997, \mn@doi [The Astrophysical Journal]
  {10.1086/303692}, 477, 118

\bibitem[\protect\citeauthoryear{{Zwaan} et~al.,}{{Zwaan}
  et~al.}{2004}]{Zwaan2004}
{Zwaan} M.~A.,  et~al., 2004, \mn@doi [\mnras]
  {10.1111/j.1365-2966.2004.07782.x}, \href
  {https://ui.adsabs.harvard.edu/abs/2004MNRAS.350.1210Z} {350, 1210}

\bibitem[\protect\citeauthoryear{{van~Eymeren}, {J{\"u}tte}, {Jog}, {Stein}  \&
  {Dettmar}}{{van~Eymeren} et~al.}{2011a}]{vanEymeren2011a}
{van~Eymeren} J.,  {J{\"u}tte} E.,  {Jog} C.~J.,  {Stein} Y.,   {Dettmar}
  R.-J.,  2011a, \mn@doi [\aap] {10.1051/0004-6361/201016177}, \href
  {http://adsabs.harvard.edu/abs/2011A%26A...530A..29V} {530, A29}

\bibitem[\protect\citeauthoryear{{van~Eymeren}, {J{\"u}tte}, {Jog}, {Stein}  \&
  {Dettmar}}{{van~Eymeren} et~al.}{2011b}]{vanEymeren2011b}
{van~Eymeren} J.,  {J{\"u}tte} E.,  {Jog} C.~J.,  {Stein} Y.,   {Dettmar}
  R.-J.,  2011b, \mn@doi [\aap] {10.1051/0004-6361/201016178}, \href
  {http://adsabs.harvard.edu/abs/2011A%26A...530A..30V} {530, A30}

\makeatother
\end{thebibliography}



\appendix

\section{SAMPLE HIPASS SPECTRA}
\label{appendix:sample_spec}

\begin{figure*}
    \centering
    \includegraphics[width=16cm]{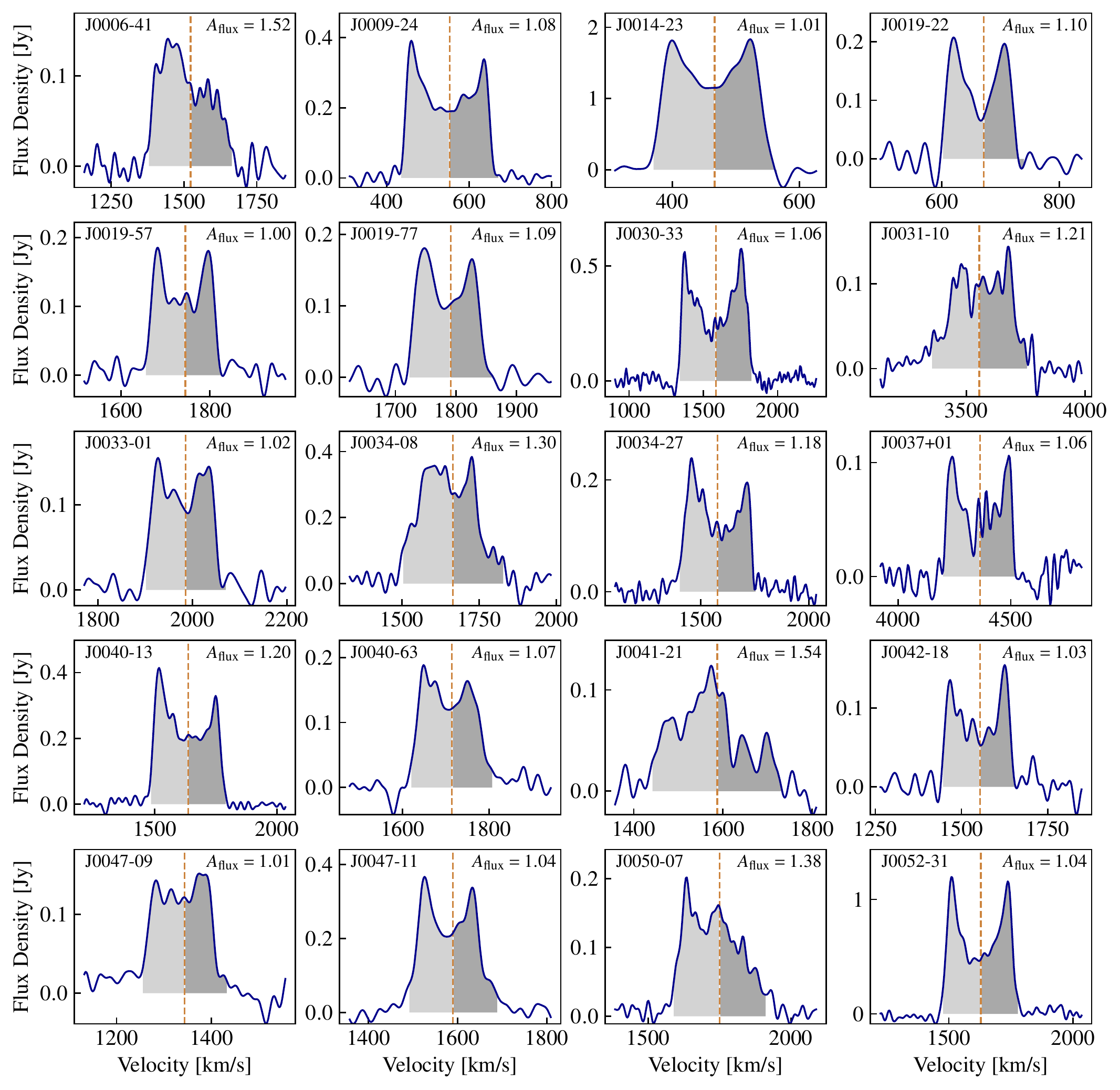}
	    \caption{Example HIPASS spectra with the flux in the lower and upper halves of the spectrum shaded (light and dark grey, respectively). The systemic velocity, $V_{\mathrm{sys,w20}}$, defined as the midpoint of the $w_{20}$ line width is indicated by the vertical dashed orange line. The calculated flux asymmetry ratio, $A_{\mathrm{flux}}$, is shown in the top right corner of each panel.}
	\label{fig:hipass_example_spectra}
\end{figure*}

In Figure~\ref{fig:hipass_example_spectra} we show a sample of 20 HIPASS spectra and their corresponding flux asymmetry ratios, $A_{\mathrm{flux}}$. The dashed orange line indicates the systemic velocity ($V_{\mathrm{sys,w20}}$) and the light and dark shaded regions indicate the integrated flux in the lower and upper halves of the spectrum ($I_1$ and $I_2$ in Equation~\ref{equ:hipass_flux_asym}, respectively) as in Figure~\ref{fig:example_spectrum}.

\bsp	
\label{lastpage}
\end{document}